\documentclass[a4paper,fleqn,usenatbib]{mnras}

\usepackage{newtxtext,newtxmath}

\usepackage[T1]{fontenc}
\usepackage{ae,aecompl}

\usepackage{graphicx}	
\usepackage{amsmath}	
\usepackage{amssymb}	

\title[Broad-band emission from engine-powered supernovae]
{Broad-band emission properties of central engine powered supernova ejecta interacting with a circumstellar medium}

\author[A. Suzuki and K. Maeda]
{Akihiro Suzuki,$^{1}$\thanks{E-mail: akihiro.suzuki@nao.ac.jp (AS), current address: National Astronomical Observatory of Japan, 2-21-1 Osawa, Mitaka, Tokyo 181-8588, Japan}
and Keiichi Maeda$^{2}$
\\
$^{1}$Yukawa Institute for Theoretical Physics, Kyoto University, Kitashirakawa-Oiwake-cho, Sakyo-ku, Kyoto, 606-8502, Japan\\
$^{2}$Department of Astronomy, Kyoto University, Kitashirakawa-Oiwake-cho, Sakyo-ku, Kyoto, 606-8502, Japan\\
}

\date{Accepted XXX. Received YYY; in original form ZZZ}

\pubyear{2017}

\begin{document}
\label{firstpage}
\pagerange{\pageref{firstpage}--\pageref{lastpage}}
\maketitle

\begin{abstract}
We investigate broad-band emission from supernova ejecta powered by a relativistic wind from a central compact object. 
A recent two-dimensional hydrodynamic simulation studying the dynamical evolution of supernova ejecta with a central energy source has revealed that outermost layers of the ejecta are accelerated to mildly relativistic velocities because of the breakout of a hot bubble driven by the energy injection. 
The outermost layers decelerate as they sweep a circumstellar medium surrounding the ejecta, leading to the formation of the forward and reverse shocks propagating in the circumstellar medium and the ejecta. 
While the ejecta continue to release the internal energy as thermal emission from the photosphere, the energy dissipation at the forward and reverse shock fronts gives rise to non-thermal emission. 
We calculate light curves and spectral energy distributions of thermal and non-thermal emission from central engine powered supernova ejecta embedded in a steady stellar wind with typical mass loss rates for massive stars. 
The light curves are compared with currently available radio and X-ray observations of hydrogen-poor superluminous supernovae, as well as the two well-studied broad-lined Ic supernovae, 1998bw and 2009bb, which exhibit bright radio emission indicating central engine activities. 
We point out that upper limits on radio luminosities of nearby superluminous supernovae may indicate the injected energy is mainly converted to thermal radiation rather than creating mildly relativistic flows owing to photon diffusion time scales comparable to the injection time scale. 
\end{abstract}



\section{INTRODUCTION\label{intro}}
Modern unbiased transient surveys have revolutionized our understanding of various explosive phenomena in the Universe. 
One of the remarkable results is the discovery of a special class of supernovae (SNe) characterized by their high luminosities ($10$--$100$ times higher than those of normal core-collapse SNe), which are now called superluminous supernovae (SLSNe; see \citealt{2012Sci...337..927G} for a review). 
Although their volumetric rate is extremely small ($<0.1\%$ of normal core-collapse SNe) \citep[e.g.][]{2013MNRAS.431..912Q,2015MNRAS.448.1206M,2017MNRAS.464.3568P}, they could be detectable at high-z galaxies thanks to their extreme luminosities \citep{2012MNRAS.422.2675T,2013MNRAS.435.2483T,2014ApJ...796...87I}. 
SLSNe are classified into a couple of subcategories based on the presence or absence of hydrogen features in their spectra. 
SLSNe without any hydrogen feature are called hydrogen-poor or type-I SLSNe (hereafter SLSNe-I) and suggested to be explosions of massive stars without hydrogen and helium envelopes \citep[e.g.][]{2007ApJ...668L..99Q,2009ApJ...690.1358B,2010ApJ...724L..16P,2011Natur.474..487Q,2011ApJ...743..114C}. 
Observations of individual SLSNe-I and their host galaxies suggest that SLSNe-I are likely produced by massive stars born in dwarf galaxies with low metallicities and high specific star formation rates \citep{2011ApJ...727...15N,2014ApJ...787..138L,2015MNRAS.449..917L,2015MNRAS.451L..65T,2016MNRAS.458...84A,2017MNRAS.470.3566C,2016ApJ...830...13P,2018MNRAS.473.1258S}.

Despite intensive observational and theoretical studies on SLSNe-I, the energy source of their bright emission is still poorly understood. 
Currently, three different scenarios have been proposed, (1) core-collapse SNe interacting with massive hydrogen-poor circumstellar matter \citep{2011ApJ...729L...6C,2012ApJ...757..178G,2013MNRAS.428.1020M}, (2) pair-instability SNe \citep{1967PhRvL..18..379B,1967ApJ...148..803R,2002ApJ...567..532H,2007Natur.450..390W,2009Natur.462..624G}, and (3) central engine powered SNe (\citealt{2010ApJ...717..245K,2010ApJ...719L.204W}; see also \citealt{1971ApJ...164L..95O,1976SvA....19..554S,2007ApJ...666.1069M}). 
The traditional and widely used way to asses the existing scenarios of the energy source is to examine whether light curves of SLSNe can successfully be explained in the framework of these scenarios. 
Supernova light curves are well explained by diffusion of thermal photons in freely expanding spherical ejecta \citep{1980ApJ...237..541A,1982ApJ...253..785A,1996snih.book.....A}. 
Therefore, the timescale of the luminosity evolution can be a key to constraining properties of exploding stars and the energy source. 
One-zone light curve models with multiple energy supplies, i.e., radioactive decay, CSM interaction, and central engine, have been formulated \citep[e.g.][]{2012ApJ...746..121C} and applied for observed light curves of SLSNe and other extraordinary SNe \citep{2013ApJ...770..128I,2013ApJ...773...76C,2015MNRAS.452.3869N,2015ApJ...799..107W,2015ApJ...807..147W,2016ApJ...821...22W,2017ApJ...837..128W,2017ApJ...850...55N}. 
Among the three scenarios, the pair-instability SN scenario requires extremely large nickel and ejecta masses, indicating slow light curve evolution. 
This is in tension with some SLSNe showing rapid evolution \citep{2013Natur.502..346N}. 
However, distinguishing these scenarios solely from IR-optical photometric observations of SLSNe is generally difficult because of a number of adjustable parameters in theoretical light curve models. 

Spectroscopic observations have also been conducted and revealed that early spectra of SLSNe-I exhibit a blue continuum and absorption features by highly-ionized oxygen \citep{2007ApJ...668L..99Q,2010ApJ...724L..16P,2011Natur.474..487Q,2013ApJ...779...98H}. 
Numerical investigations on the spectral formation in SLSNe-I have also been attempted \citep{2012MNRAS.426L..76D,2016MNRAS.458.3455M}. 
Spectra of SLSNe-I at various epochs are available particularly for well-observed nearby events, such as SN 2007bi, PTF09cnd, 2010gx, PS1-11ap, PTF12dam, LSQ14an, LSQ14mo, and SN 2015bn \citep[e.g.][]{2009Natur.462..624G,2010ApJ...724L..16P,2011Natur.474..487Q,2013Natur.502..346N,2014MNRAS.437..656M,2015ApJ...807L..18N,2015MNRAS.452.1567C,2015ApJ...815L..10L,2017MNRAS.468.4642I,2017A&A...602A...9C}. 
\cite{2017ApJ...845...85L} compared spectra of SLSNe-I and stripped-envelope CCSNe in a systematic way by using the Markov Chain Monte Carlo (MCMC)-based spectral fitting method developed by 
\cite{2016ApJ...827...90L} and \cite{2016ApJ...832..108M}. 
They found that the average photospheric velocity of SLSNe-I implied by FeII absorption lines ($\sim15000$ km s$^{-1}$ around 10 days after the peak) is higher than normal type Ic SNe ($\sim 7000$ km s$^{-1}$) and similar to type Ic SNe characterized by broad absorption features (SNe Ic-BL). 
Recent observations of the SLSN-I 2015bn at $z=0.1136$ also showed that the nebular spectra at later epochs were remarkably similar to those of SNe Ic-BL \citep{2016ApJ...828L..18N,2017ApJ...835...13J}. 
These findings may indicate a link between the two extraordinary classes of SNe, SLSNe-I and SNe Ic-BL.  
However, theoretical understanding of characteristic spectral features associated with different scenarios has still been limited, which makes it difficult by optical spectra to distinguish different scenarios. 

Furthermore, an SLSN-like bump is found in the afterglow light curve of the ultra-long gamma-ray burst (GRB) 111209A and named SN 2011kl \citep{2015Natur.523..189G,2016arXiv160606791K}. 
These observations may further support the scenario that SLSNe-I and SNe Ic-BL (or associated GRBs) are powered by the same engine. 
Actually, a fast rotating magnetized neutron star, which is the most popular power source for the central engine scenario of SLSNe-I, have also been considered as a potential central engine for GRBs \citep{1992Natur.357..472U,2000ApJ...537..810W,2004ApJ...611..380T,2008MNRAS.383L..25B,2009MNRAS.396.2038B,2011MNRAS.413.2031M}. 
More recently, roles of a mili-second magnetized neutron star newly born in supernova ejecta are also paid a great attention in the context of fast radio bursts (FRBs). 
The recently realized localization of the repeating FRB 121102 and the associated persistent radio source have stimulated intense discussion on its progenitor and emission mechanism \citep{2017Natur.541...58C,2017ApJ...834L...8M,2017ApJ...834L...7T}. 
The similarity of the host galaxy of the FRB and those of SLSNe may indicate the possible FRB-SLSN or FRB-SLSN-GRB association \citep{2017ApJ...834L...7T,2017ApJ...841...14M,2017ApJ...843...84N}.

Another potential way to distinguishing energy sources of SLSNe-I is to identify emission signatures across a wide energy range from radio to gamma-rays. 
Radio waves and high energy photons from young CCSNe are usually attributed to emission from non-thermal electrons produced by blast waves driven by supernova ejecta. 
Such non-thermal emission is also naturally expected for SLSNe-I.  
Multi-wavelength observations of SLSNe-I have been conducted particularly for nearby events, such as SN 2015bn \citep{2016ApJ...826...39N}, Gaia16apd \citep{2018ApJ...856...56C}, and SN 2017egm (also known as Gaia17biu) \citep{2018ApJ...853...57B,2017ApJ...845L...8N}. 
Systematic searches for X-ray emission from SLSNe-I have been carried out and compiled by \cite{2013ApJ...771..136L} and \cite{2017arXiv170405865M}. Currently, possible detections of X-ray sources whose locations are consistent with SCP 06F6 \citep{2009ApJ...697L.129G} and PTF12dam \citep{2017arXiv170405865M} have been reported. 
However, the origin of the X-ray emission is still debated. 
Some SLSNe-I, e.g., Gaia 16apd, exhibit a significant UV excess, which should also be a key to revealing the energy source \citep{2017ApJ...840...57Y,2017ApJ...835L...8N,2017MNRAS.469.1246K,2017ApJ...845L...2T}. 
Furthermore, a systematic search for gamma-ray emission associated with SLSNe has been conducted by \cite{2018A&A...611A..45R}, although they obtained only an upper limit for the gamma-ray luminosity by assuming a photon spectrum of $\nu^{-2}$.

If an SLSNe-I is powered by a relativistic wind from a fast-rotating magnetized neutron star in an analogy to Galactic pulsar wind nebulae, electron-positron pairs would be copiously produced in the downstream of the shock wave terminating the wind. 
These high energy particles with non-thermal energy spectra can serve as an ionizing photon source for the supernova ejecta surrounding the neutron star. 
Thus, the presence of a nascent neutron star at the centre of the expanding supernova ejecta could be probed by the ionization structure of the ejecta and/or radio, X-ray, and gamma-ray emission \citep{2013MNRAS.432.3228K,2014MNRAS.437..703M,2015ApJ...805...82M,2016MNRAS.461.1498M,2016ApJ...818...94K}. 
\cite{2017ApJ...841...14M} considered radio emission associated with SLSNe in the context of the FRB-SLSNe connection and pointed out that the quiescent radio source found in the host galaxy of FRB 121102 could be an SLSN remnant having produced a magnetar. 
They considered radio emission from the pulsar wind nebula and the forward shock driven by the supernova ejecta. 
More recently, \cite{2018MNRAS.474..573O} present similar calculations based on the model developed by \cite{2015ApJ...805...82M} and \cite{2016ApJ...818...94K}.

Light curve modellings of central engine powered supernova ejecta in a wide range of wavelengths would greatly help us understand the emission mechanism of SLSNe-I and ultimately unveil their enigmatic origin. 
Especially, radio and X-ray emission from SNe can probe the density and velocity structure of supernova ejecta and also is expected to play a role in distinguishing the existing scenarios of SLSNe-I. 
Recent numerical studies on supernova ejecta with central energy injection have claimed that multi-dimensional effects are important in determining the density structure of the ejecta. 
Multi-dimensional effects of the central energy injection into supernova ejecta, such as mixing and breakout of a hot bubble, have been considered by several authors \citep{2003ApJ...589..871A,2011MNRAS.411.2054L}. 
However, it is only recently that such effects are investigated by using hydrodynamic simulations in the context of SLSNe \citep{2016ApJ...832...73C,2017MNRAS.466.2633S,2017ApJ...845..139B}. 
The two-dimensional hydrodynamics simulation of the interaction between supernova ejecta and a relativistic wind from the central engine presented by \cite{2017MNRAS.466.2633S} revealed that the interaction leads to the creation of a hot bubble at the centre of the ejecta, which eventually blows out the whole ejecta and accelerates outermost layers to mildly relativistic speeds. 
These features are very different from the conventional ejecta structure applied to normal CCSNe. 
The mildly relativistic component of the supernova ejecta colliding with the CSM potentially produces bright non-thermal emission, which can serve as a signature of the hot bubble breakout.

In this paper, we present a theoretical model for photospheric, synchrotron, and inverse Compton emission from central engine powered supernova ejecta and the accompanying blast wave in the CSM. 
The thermal and non-thermal emission are calculated by a method commonly used in supernova studies, while we adopt the ejecta profile indicated by the hydrodynamic simulation. 
The dynamical evolution of the supernova ejecta expanding into the CSM is treated by the method developed by \cite{2017ApJ...834...32S}, who considered the hydrodynamic collision of trans-relativistic spherical ejecta with a steady stellar wind. 
In Section \ref{sec:dynamics}, we describe our model for the dynamical evolution and photospheric emission of freely expanding supernova ejecta with a power-law density profile. 
In Section \ref{sec:non_thermal}, details of the non-thermal emission model are presented. 
In Section \ref{sec:results}, we compare our theoretical light curves with currently available observations of SLSNe-I and the two SNe Ic-BL 1998bw \citep{1998Natur.395..663K,1998Natur.395..670G} and 2009bb  \citep{2010Natur.463..513S}, which are characterized by bright radio emission indicating the presence of a central engine activity. 
Finally, we conclude this paper in Section \ref{sec:conclusion}

\section{Dynamical evolution and thermal emission of supernova ejecta}\label{sec:dynamics}
In this section, we describe our model for the dynamical evolution of supernova ejecta powered by a central engine. 
\subsection{Energy injection from the central engine}
Our previous simulation \citep{2017MNRAS.466.2633S} is based on the central-engine scenario for SLSNe \citep{2010ApJ...717..245K} and has assumed energy injection at a constant rate around the centre. 
On the other hand, the most popular model for central-engine powered supernovae adopts mili-second magnetar spin down as the primary power source. 
The energy injection rate is usually assumed to be proportional to $(1+t/t_\mathrm{sd})^{-s}$, where $t_\mathrm{sd}$ is the spin down time of the magnetar and $s$ is an exponent (hereafter, $s=2$). 
Therefore, while the energy injection rate is constant well before the characteristic spin down time $t_\mathrm{sd}$, it decays in a power-law fashion at $t\gg t_\mathrm{sd}$. 
In order to incorporate the multi-dimensional picture revealed by the numerical simulation into calculations of thermal and non-thermal emission powered by the central engine, we assume that the structure of the ejecta has been fixed after the total energy of the ejecta reaches $E_\mathrm{ej}=10^{52}$ erg at $t=t_\mathrm{sd}$. 
The radial density and velocity profiles of the ejecta are assumed to be those derived by \cite{2017MNRAS.466.2633S}, which are reviewed in the next subsection. 
The normalization of the spin down energy deposition rate $L_\mathrm{sd}$ is determined so that the deposited energy reaches $E_\mathrm{ej}$ at $t=t_\mathrm{sd}$. 
Thus, the spin down rate is expressed as follows,
\begin{equation}
L_\mathrm{sd}(t)=\frac{2E_\mathrm{ej}}{t_\mathrm{sd}}(1+t/t_\mathrm{sd})^{-2}
\label{eq:L_sd}
\end{equation}
With this normalization, the total deposited energy yields $2E_\mathrm{ej}$. 
The energy deposited at $t>t_\mathrm{sd}$ serves as a power source for thermal emission from the ejecta. 
Although a fraction of the energy may be used to accelerate the ejecta, it is smaller than the total energy of the ejecta and thus unlikely to significantly affect the subsequent dynamical evolution of the ejecta. 
This treatment makes the dynamical model not fully self-consistent. 
Nevertheless, important aspects of the dynamical evolution of the ejecta are certainly captured. 

In this work, we are interested in thermal and non-thermal emission from the ejecta having experienced the hot bubble breakout. 
We start the calculation of the emission at the initial time $t_\mathrm{i}=t_\mathrm{sd}$. 
The spin down time, which is now equal to the initial time of the calculation, is a free parameter specifying the timescale of the central energy injection. 

\subsection{Supernova ejecta with central energy injection}
We review the dynamical evolution of supernova ejecta powered by the central energy injection at a constant rate and how the subsequent energy redistribution throughout the ejecta shapes the density and velocity structure of the ejecta. 

\subsubsection{Powering supernova ejecta}\label{sec:powering_sn_ejecta}
\cite{2017MNRAS.466.2633S} have considered the dynamical evolution of supernova ejecta powered by a relativistic wind injected from a central engine at a constant energy injection rate. 
The ejecta are assumed to be expanding in a homologous way, i.e., the radial velocity $v$ of a layer is its radius divided by the elapsed time, $v=r/t$. 
A widely used broken power-law model \citep{1989ApJ...341..867C}, where the density is proportional to the radial velocity, $\rho\propto v^{-\delta}$ for inner ejecta and $\rho\propto v^{-m}$ for outer ejecta, is employed for the density profile of the supernova ejecta. 
The inner density gradient should be shallow, $\delta<3$, so that the mass of the ejecta should not diverge at the centre. 
On the other hand, the outer density gradient is usually assumed to be steep, with a typical index of $m\sim 10$. 
The numerical simulation adopted $\delta=1$ and $m=10$. 

The important parameters characterizing the dynamical evolution of the ejecta are the original kinetic energy of the ejecta $E_\mathrm{sn}$ and the energy injection rate $\dot{E}_\mathrm{in}$. 
These two parameters give the characteristic timescale $t_\mathrm{c}=E_\mathrm{sn}/\dot{E}_\mathrm{in}$, at which the injected energy reaches the original kinetic energy. 
The dynamical evolution can be scaled by this critical timescale. 
In other words, we can apply the following scenario for different energy injection rates by rescaling the time $t$. 

\begin{figure*}
\begin{center}
\includegraphics[scale=0.8,bb=0 0 565 225]{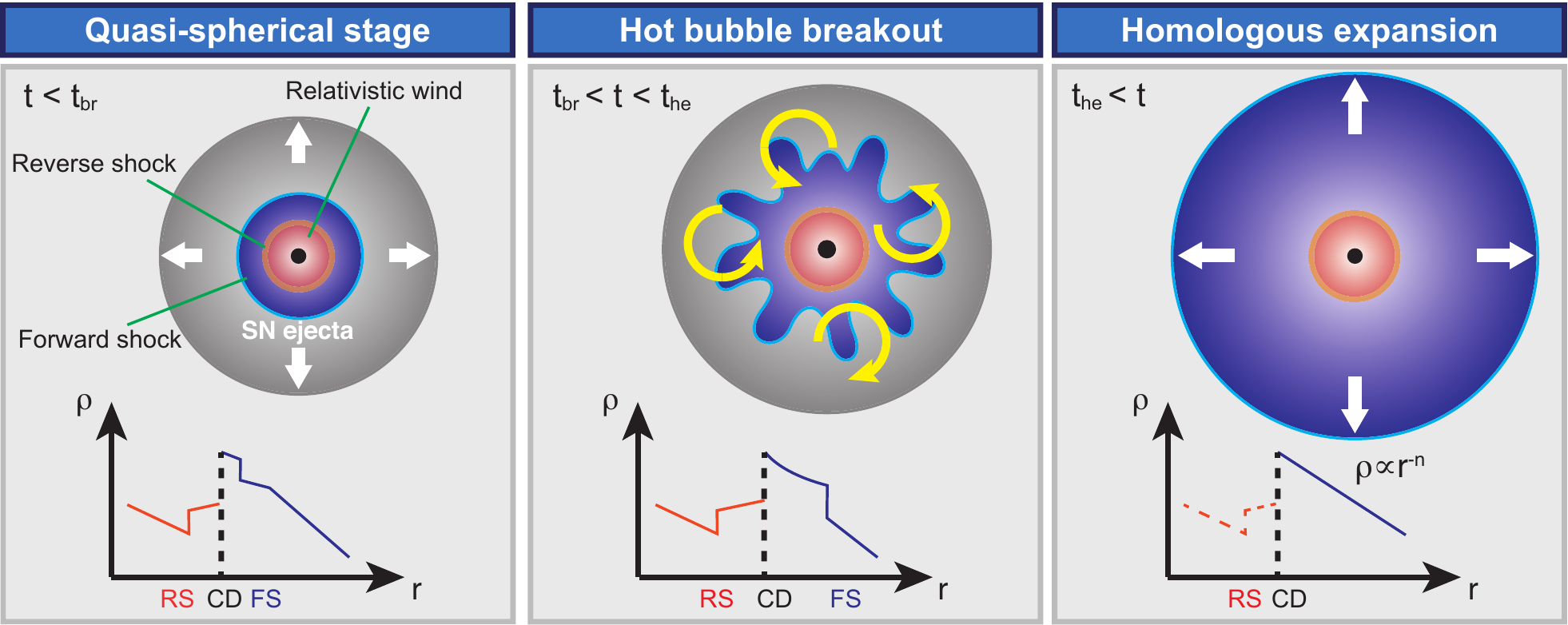}
\caption{Schematic views of the dynamical evolution of supernova ejecta with a relativistic wind. 
The three stages, (1) quasi-spherical, (2) hot bubble breakout, and (3) homologous expansion stages, are depicted from left to right. 
The reverse shock, the contact discontinuity, and the forward shock are denoted by RS, CD, and FS in the density profiles. }
\label{fig:schematic}
\end{center}
\end{figure*}

The numerical simulation revealed that the dynamical evolution of supernova ejecta with an embedded relativistic wind can be divided into the following three stages (see the schematic representation in Figure \ref{fig:schematic}): 
(1) The relativistic wind injected around the centre first creates a quasi-spherical, geometrically thin shell composed of the shocked wind and ejecta (quasi-spherical stage). 
In this stage, the shocked gas forms a quasi-spherical hot bubble well confined by the ram pressure of the ejecta. 
The dynamical evolution of the shell in this stage is described by a self-similar solution and the radius of the shell evolves as $t^{\alpha}$, where $\alpha=(6-\delta)/(5-\delta)$ \citep{1982ApJ...258..790C,1998ApJ...499..282J,2005ApJ...619..839C}. 
(2) When the forward shock propagating in the ejecta reaches a layer above which the density gradient is steep, the ram pressure of the ejecta no longer confines the hot bubble. 
As a result, the steep density gradient efficiently accelerates the forward shock and the whole ejecta are gradually overwhelmed by the shocked gas (hot bubble breakout). 
This transition happens at $t_\mathrm{br}=f_\mathrm{br}t_\mathrm{c}$, when the total amount of the energy injected from the central engine exceeds a threshold value $f_\mathrm{br}E_\mathrm{sn}$. 
The factor $f_\mathrm{br}$ depends on the structure of the ejecta. 
We assume $f_\mathrm{br}=5$ (\citealt{2017MNRAS.466.2633S}; see also \citealt{2017ApJ...845..139B}).
(3) After the emergence of the forward shock from the outermost layer of the ejecta, the energy of the ejecta is gradually redistributed and the ejecta approach the homologous expansion stage. 
The density structure in this stage is well represented by a power-law function of the radial velocity with an exponent $-6$ \citep{2017MNRAS.466.2633S}.

\subsubsection{Homologous expansion of supernova ejecta}\label{sec:ejecta}
We focus on the evolution of the supernova ejecta after the power-law density structure is realized ($t>t_\mathrm{sd}$). 
The velocity distribution is again represented by 
\begin{equation}
v(t,r)=\left\{\begin{array}{ccl}
r/t&\mathrm{for}&r\leq v_\mathrm{max}t,\\
0&\mathrm{for}&v_\mathrm{max}t<r,
\end{array}\right.
\label{eq:velocity_profile}
\end{equation}
with a maximum velocity $v_\mathrm{max}$. 
We assume that the density profile of the freely expanding ejecta is described by a power-law function of the four-velocity with an exponent $-n$,
\begin{equation}
\rho(t,r)=\left\{
\begin{array}{cl}
\rho_0\left(\frac{t}{t_\mathrm{i}}\right)^{-3}\left(\frac{\Gamma v}{\Gamma_\mathrm{max}v_\mathrm{max}}\right)^{-n}
&\mathrm{for}\ \ \ v_\mathrm{min}t\leq r\leq v_\mathrm{max}t,\\
0&\mathrm{otherwise},
\end{array}
\right.
\label{eq:density_profile}
\end{equation}
where the Lorentz factor is given by
\begin{equation}
\Gamma=\frac{1}{\sqrt{1-(v/c)^2}}.
\end{equation}
In this study, we use a fixed value for the maximum four-velocity, $\Gamma_\mathrm{max}v_\mathrm{max}=c$, following our previous simulation \citep{2017MNRAS.466.2633S}. 
The outermost layer travelling at the maximum velocity has been transparent to optical photons at the time of creation. 
Thus, optical emission would not be affected by the adopted maximum velocity. 
Furthermore, the layer will soon be swept by the reverse shock, making the subsequent dynamical evolution and non-thermal emission from the shocked gas insensitive to the assumed value \citep{2017ApJ...834...32S}. 
Our previous study showed that the angle-averaged density structure of the supernova ejecta is well represented by a power-law profile with an exponent $n=6$ (see, Section \ref{sec:powering_sn_ejecta}). 
We use $n=6$ as our fiducial value and examine how different values affect the non-thermal emission. 
The normalization constant $\rho_0$ and the minimum velocity $v_\mathrm{min}$ are determined for a given set of the ejecta mass and energy, $M_\mathrm{ej}$ and $E_\mathrm{ej}$, as follows,
\begin{equation}
M_\mathrm{ej}=4\pi \rho_0(ct_\mathrm{i})^3\int_{v_\mathrm{min}}^{v_\mathrm{max}}
\Gamma\left(\frac{\Gamma v}{\Gamma_\mathrm{max}v_\mathrm{max}}\right)^{-n}v^2dv,
\end{equation}
and
\begin{equation}
E_\mathrm{ej}=4\pi \rho_0c^5t_\mathrm{i}^3\int_{v_\mathrm{min}}^{v_\mathrm{max}}
\Gamma(\Gamma-1)\left(\frac{\Gamma v}{\Gamma_\mathrm{max}v_\mathrm{max}}\right)^{-n}v^2dv.
\end{equation}
We assume $M_\mathrm{ej}=10M_\odot$ and $E_\mathrm{ej}=10^{52}$ erg in order to imitate the freely expanding ejecta realized in our previous numerical simulation  \citep{2017MNRAS.466.2633S}.

\subsection{Photospheric emission}
The supernova ejecta powered by the central energy injection give rise to bright thermal emission \citep{2010ApJ...717..245K}. 
We consider thermal photons diffusing out from the ejecta after $t=t_\mathrm{i}$. 
The photospheric radius $R_\mathrm{ph}$ at time $t$ can be calculated in the following way. 
The optical depth for a ray radially extending from a given radius $r$ to the outermost radius of the ejecta is calculated by
\begin{equation}
\tau(r,t)=\int_r^{v_\mathrm{max}t}\kappa\rho(t,r') dr',
\label{eq:tau}
\end{equation}
where $\kappa$ is the opacity for thermal photons and set to be $\kappa=0.1$ cm$^2$ g$^{-1}$. 
Here we have ignored the motion of the ejecta while the ray is travelling. 
In addition, the outermost layers of the ejecta would be swept by the reverse shock and thus the density structure is modified. 
We also ignore the modification of the density structure for simplicity. 
The photospheric radius at $t$ is determined so that the optical depth is equal to unity, $\tau(R_\mathrm{ph},t)=1$. 
We particularly denote the photospheric radius at $t=t_\mathrm{i}$ by $R_\mathrm{i}$. 

We calculate the photospheric emission from the ejecta being powered by the continuous energy injection at the centre. 
We basically use the Arnett's solution for photon diffusion in freely expanding spherical ejecta \citep{1980ApJ...237..541A,1982ApJ...253..785A}. 
The bolometric luminosity of the photospheric emission from the ejecta with energy input $L_\mathrm{in}(t)$ is given by
\begin{eqnarray}
L_\mathrm{ph}(t)&=&\frac{2}{t_\mathrm{d}}e^{-t(t+2t_\mathrm{h})/t_\mathrm{d}^2}
\int^t_{t_\mathrm{i}}e^{t'(t+2t_\mathrm{h})/t_\mathrm{d}^2}L_\mathrm{in}(t')
\left(\frac{t_\mathrm{h}}{t_\mathrm{d}}+\frac{t'}{t_\mathrm{d}}\right)dt'
\nonumber
\\&&
+\frac{E_\mathrm{th,0}}{t_0}e^{-t(t+2t_\mathrm{h})/t_\mathrm{d}^2},
\end{eqnarray}
where the timescales $t_0$, $t_\mathrm{h}$, and $t_\mathrm{d}$ are given by
\begin{equation}
t_\mathrm{0}=\frac{\kappa M_\mathrm{ej}}{\beta cR_\mathrm{i}},
\end{equation}
\begin{equation}
t_\mathrm{h}=\frac{R_\mathrm{i}}{v(R_\mathrm{i})},
\end{equation}
and
\begin{equation}
t_\mathrm{d}=\sqrt{2t_0t_\mathrm{h}},
\end{equation}
\citep{2012ApJ...746..121C,2013ApJ...770..128I}. 
Here $E_\mathrm{th,0}$ is the initial thermal energy and $v(R_\mathrm{i})$ is the radial velocity at the photosphere $r=R_\mathrm{i}$, both given at $t=t_\mathrm{i}$. 
The initial thermal energy can be obtained from the dynamical model. 
The thermal energy of the ejecta in the quasi-spherical stage almost linearly increases with time \citep{2017MNRAS.466.2633S}. 
The value at the end of the increase is given by
\begin{equation}
E_\mathrm{th}=\frac{2-\gamma}{1+3\alpha(\gamma-1)}E_\mathrm{ej},
\label{eq:E_th}
\end{equation}
where $\gamma=4/3$ is the adiabatic index. 
The non-dimensional constant $\beta$ depending on the density structure is set to be a commonly used value $\beta=13.8$ \citep{1980ApJ...237..541A,1982ApJ...253..785A}. 

We assume the energy injection at the rate given by Equation (\ref{eq:L_sd}), where the spin down time $t_\mathrm{sd}$ is one of our input parameters, the effects of which are to be examined in this paper. 
The energy input into the ejecta is given by
\begin{equation}
L_\mathrm{in}(t)=L_\mathrm{sd}(t)(1-e^{-\tau_\gamma}),
\end{equation}
where the last factor takes into account the leakage of the injected energy from the ejecta as gamma-rays \citep{2015ApJ...799..107W}. 
We calculate the optical depth by the following integration,
\begin{equation}
\tau_\gamma=\kappa_\gamma\int^{v_\mathrm{max}t}_{v_\mathrm{min}t}\rho(t,r)dr,
\end{equation}
which evolves as $\tau_\gamma\propto t^{-2}$. 

The gamma-ray opacity $\kappa_\gamma$ should depend on the frequency of gamma-rays and the effective value may possibly be dependent  the three-dimensional density distribution of the supernova ejecta.
Therefore, the value is highly uncertain. 
Theoretical calculations by \cite{2013MNRAS.432.3228K}, who assumed simplified spherical ejecta, suggest that the gamma-ray opacity could be of the order of $\sim 0.1$ cm$^2$ g$^{-1}$ for photons with $h\nu\simeq 100$ keV because of Compton scattering and $\sim0.01$ cm$^2$ g$^{-1}$ for photons with $h\nu>10$ MeV because of pair production. 
We should note that the effective opacity may be lower than these values when we take into account patchy density structure. 
Recently, several authors have incorporated the gamma-ray leakage effect into their light curve fitting models and tried to constrain the gamma-ray opacity. 
\cite{2017ApJ...842...26L} fitted light curves of 19 SLSNe-I by their light curve model and analyzed the results by an MCMC approach. 
They reported that the best-fit value of the gamma-ray opacity ranges from $\kappa_\gamma\simeq 0.01$ cm$^{2}$ g$^{-1}$ to $\kappa_\gamma\simeq 0.82$ cm$^{2}$ g$^{-1}$. 
\cite{2017ApJ...850...55N} systematically studied multi-colour light curves of 38 SLSNe-I.  
For example, their analysis on SN 2015bn inferred $\kappa_\gamma\simeq 0.01$ gm$^{2}$ g$^{-1}$. 
Other SLSNe with well-covered late-time evolutions also showed small values, indicating significant gamma-ray leakage especially at later epochs. 
Keeping in mind that the value should be treated with caution, we adopt a constant value of $\kappa_\gamma=0.01$ cm$^2$ g$^{-1}$. 

Finally, we determine the temperature of the photospheric emission. 
Determining the colour temperature of the photospheric emission requires sophisticated treatments of radiative transfer, the ionization states of different layers of the ejecta, and numerous line opacities contributing to the thermal balance of the ejecta. 
For simplicity, we assume that the spectrum of the emission is well represented by a Planck function. 
Thus, we estimate the effective temperature $T_\mathrm{eff}$ of the photospheric emission from the photospheric radius and the bolometric luminosity given above,
\begin{equation}
L_\mathrm{ph}=4\pi R_\mathrm{ph}^2\sigma_\mathrm{SB}T_\mathrm{eff}^4,
\end{equation}
where $\sigma_\mathrm{SB}$ is the Stefan-Boltzmann constant. 

\subsection{Ejecta-CSM interaction}
\cite{2017ApJ...834...32S} considered the hydrodynamical interaction between spherical supernova ejecta travelling at mildly relativistic speeds and a steady wind with a mass-loss rate $\dot{M}$ and a wind velocity $v_\mathrm{w}$,
\begin{equation}
\rho_\mathrm{csm}=\frac{\dot{M}}{4\pi v_\mathrm{w} r^2}\equiv Ar^{-2}
\end{equation}
where $A$ is a free parameter specifying the CSM density. 
We introduce the non-dimensional parameter $A_\star=A/(5\times 10^{11}\ \mathrm{g\ cm}^{-1})$. 
In this normalization, $A_\star=1$ corresponds to a mass-loss rate of $\dot{M}= 10^{-5}\ M_\odot$ yr$^{-1}$ for a wind velocity of $10^3$ km s$^{-1}$. 
In the following, we assume CSM density parameters up to $A_\star=10$, with which the CSM is still transparent for electron scattering. 
Thus we can safely assume that the ejecta-CSM interaction does not give rise to optically thick thermal radiation significantly contributing to the optical brightness of SNe. 
We use this semi-analytic model to describe the evolution of the forward and reverse shocks developed as a result of the collision of the ejecta with the CSM. 

For a given set of the parameters, $M_\mathrm{ej}$, $E_\mathrm{ej}$, and $n$, the density and radial velocity profiles of the ejecta are specified. 
Under the assumption that the ejecta start interacting with the surrounding gas at $t=t_\mathrm{i}$, the semi-analytic model is used to calculate the shock radius, the rate of the energy dissipation via shock, and the swept mass, as a function of time for both the forward and reverse shocks. 
The rate of the energy dissipation at the shock front and the mass swept by the shock are used to specify the number and the average energy of non-thermal electrons injected into the shocked region as we describe in the next section. 

\begin{figure}
\begin{center}
\includegraphics[scale=0.5,bb=0 0 432 576]{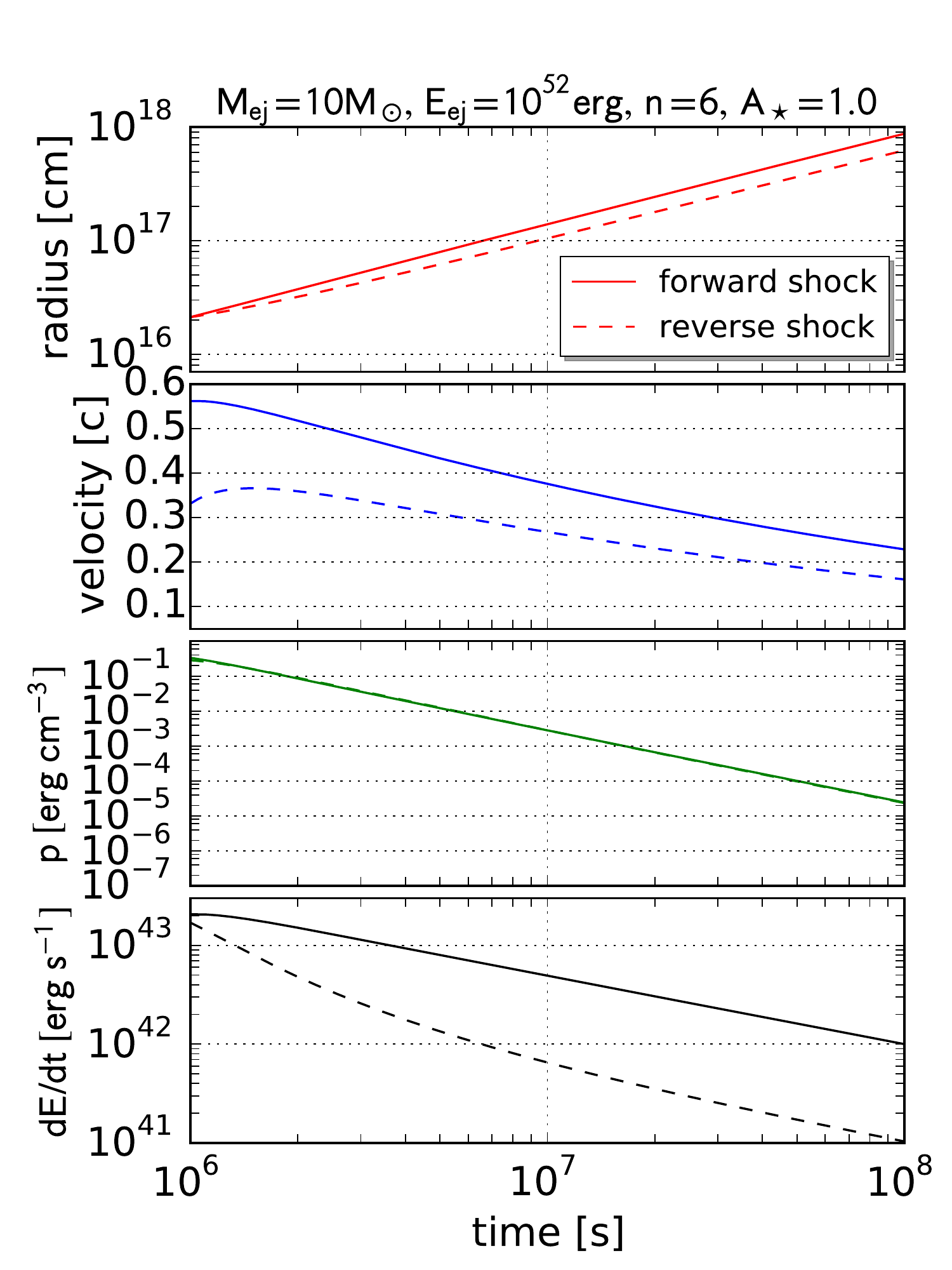}
\caption{Temporal evolutions of the shock radius (top panel), shock velocity (middle panel), and the energy dissipation rate (bottom panel) calculated by the one-zone model. 
In each panel, the solid and dashed lines correspond to the quantities at the forward and reverse shock fronts, respectively. 
The parameters specifying the ejecta and CSM models are assumed to be $M_\mathrm{ej}=10\ M_\odot$, $E_\mathrm{ej}=10^{52}$ erg, $n=6$, and $A_\star=1.0$. }
\label{fig:dynamics}
\end{center}
\end{figure}

Figure \ref{fig:dynamics} shows an example of the semi-analytic calculation with $t_\mathrm{sd}=10^6$ s. 
The temporal evolution of the shock radius, the shock velocity, the post-shock pressure, and the internal energy dissipation rate are plotted for the forward and reverse shocks. 
After the beginning of the calculation at $t=t_\mathrm{i}$($=10^6$ s), the forward and reverse shock radii steadily increase with time. 
The difference in the forward and reverse shock radii is much smaller than the shock radius, indicating that the shocked region can be described as a geometrically thin shell. 
The shock velocities decrease to $\sim 0.2c$ by the end of the calculation at $t=10^8$ s. 
The post-shock pressure at the forward and reverse shock fronts are similar because of the pressure balance across the contact discontinuity separating the shocked ejecta and CSM.

\section{Non-thermal emission model}\label{sec:non_thermal}
In this section, we describe our non-thermal emission model. 
We treat the thin shocked region under a one-box approximation and calculate the temporal evolution of the isotropic momentum distribution, $dN/dp_\mathrm{e}$, of electrons uniformly distributed in the shocked region. 
Non-thermal electrons with a power-law momentum distribution are injected through the forward and reverse shocks and then they lose their energies via radiative and adiabatic cooling. 
We consider synchrotron and inverse Compton emission as radiative cooling processes. 
From the temporal evolution of the electron distribution, we calculate the spectrum of the non-thermal emission from the shocked gas. 
\subsection{Electron injection at the shock front}
The shock dissipation at the forward and reverse shock fronts creates non-thermal electrons. 
Our treatment of the electron injection is similar to studies on non-thermal emission from CCSNe in the literature (e.g., \citealt{1998ApJ...499..810C,2006ApJ...651..381C}; see \citealt{2016arXiv161207459C} for a recent review). 
We introduce two free parameters, $\epsilon_\mathrm{e}$ and $\epsilon_\mathrm{B}$, representing the efficiencies of the non-thermal electron acceleration and magnetic field generation at the shock front. 
These values should ideally be self-consistently determined by microscopic plasma processes responsible for the energy equipartition in the shock downstream. 
However, how exactly electrons are accelerated and magnetic fields are amplified are still debated even with state-of-the-art numerical computations based on first principle approaches \citep[e.g.,][]{2008ApJ...682L...5S}. 
Thus, we fix their values to be constant. 

The energy injected into the shocked region per unit time is obtained by the semi-analytic model described in Section \ref{sec:dynamics}. 
We denote the energy injection rates at the forward and reverse shocks by $\dot{E}_\mathrm{fs}$ and $\dot{E}_\mathrm{rs}$. 
For a given set of the post-shock density $\rho$ and the internal energy density $u_\mathrm{int}$, the internal energy density $u_\mathrm{ele}$ and the number density $n_\mathrm{ele}$ of the injected non-thermal electrons are
\begin{equation}
u_\mathrm{ele}=\epsilon_\mathrm{e}u_\mathrm{int},
\end{equation}
and
\begin{equation}
n_\mathrm{ele}=\frac{Z\rho}{Am_\mathrm{u}},
\end{equation}
where $A$ and $Z$ are the mass and atomic numbers of ions predominantly composing the ejecta and $m_\mathrm{u}$ is the atomic mass unit. 
We assume $Z/A=0.5$ in the following. 
The average energy of the injected non-thermal electrons is obtained as follows,
\begin{equation}
\bar{\gamma}m_\mathrm{e}c^2=\frac{u_\mathrm{ele}}{n_\mathrm{ele}}=\frac{\epsilon_\mathrm{e}Am_\mathrm{u}u_\mathrm{int}}{Z\rho},
\label{eq:electron_gamma}
\end{equation}
where $m_\mathrm{e}$ is the electron mass. 
We assume that electrons are spontaneously accelerated by the shock passage and then obey the following simple power-law momentum distribution,
\begin{equation}
\left(\frac{d\dot{N}}{dp_\mathrm{e}}\right)_\mathrm{in}=
\left\{\begin{array}{cl}
0&\mathrm{for}\ \ p_\mathrm{min}\leq p_\mathrm{e}\leq p_\mathrm{in},\\
K(p_\mathrm{e}/p_\mathrm{in})^{-p}&\mathrm{for}\ \ p_\mathrm{in}\leq p_\mathrm{e}\leq p_\mathrm{max},\\
\end{array}\right.
\end{equation}
where $p_\mathrm{e}$ is the electron momentum and $p_\mathrm{min}$ and $p_\mathrm{max}$ are the minimum and maximum values. 
The minimum and maximum momenta are set to be $p_\mathrm{max}=10^{-3}m_\mathrm{e}c$ and $p_\mathrm{max}=10^6m_\mathrm{e}c$. 
We assume a power-law index of $p=3$, which is commonly employed to account for radio observations of stripped-envelope CCSNe \citep[e.g.,][]{2006ApJ...651..381C}. 
Therefore, electrons at the minimum injection energy $c(m_\mathrm{e}^2c^2+p_\mathrm{in}^2)^{1/2}$ carry a considerable fraction of the internal energy of electrons. 
The normalization constant $K$ and the injection momentum $p_\mathrm{in}$ characterizing the momentum distribution of the injected electrons are determined by equating the mass and energy injection rates and the following two integrals,
\begin{equation}
\int_{p_\mathrm{in}}^{p_\mathrm{max}}\left(\frac{d\dot{N}}{dp_\mathrm{e}}\right)_\mathrm{in}dp_\mathrm{e}=
\frac{\dot{E}}{\bar{\gamma}m_\mathrm{e}c^2},
\end{equation}
and
\begin{equation}
\int_{p_\mathrm{in}}^{p_\mathrm{max}}
c(m_\mathrm{e}^2c^2+p_\mathrm{e}^2)^{1/2}
\left(\frac{d\dot{N}}{dp_\mathrm{e}}\right)_\mathrm{in}dp_\mathrm{e}=
\dot{E},
\end{equation}
with $\dot{E}=\dot{E}_\mathrm{fs}$ or $\dot{E}_\mathrm{rs}$. 

\subsection{Electron momentum distribution}
The electrons injected at the shock front experience synchrotron, inverse Compton, and adiabatic cooling. 
The governing equation describing the temporal evolution of the electron momentum distribution is written as follows,
\begin{equation}
\frac{\partial }{\partial t}\left(\frac{dN}{dp_\mathrm{e}}\right)
=\frac{\partial }{\partial p_\mathrm{e}}
\left[
\left(\dot{p}_\mathrm{syn}+\dot{p}_\mathrm{ic}+\dot{p}_\mathrm{ad}\right)
\frac{dN}{dp_\mathrm{e}}
\right]
+\left(\frac{d\dot{N}}{dp_\mathrm{e}}\right)_\mathrm{in},
\label{eq:momentum_equation}
\end{equation}
where $\dot{p}_\mathrm{syn}$, $\dot{p}_\mathrm{ic}$, and $\dot{p}_\mathrm{ad}$ are the momentum loss rates by the three cooling processes. 
The synchrotron and inverse Compton momentum loss rate are related to the corresponding energy loss rates, $\dot{E}_\mathrm{syn}$ and $\dot{E}_\mathrm{ic}$, as follows, 
\begin{equation}
\dot{p}_{\{\mathrm{syn,ic}\}}=\frac{\sqrt{m_\mathrm{e}^2c^2+p_\mathrm{e}^2}}{p_\mathrm{e}c}
\dot{E}_{\{\mathrm{syn,ic}\}}.
\end{equation}
The synchrotron and inverse Compton energy loss rates are given by
\begin{equation}
\dot{E}_\mathrm{syn}=\frac{4}{3}\sigma_\mathrm{T}cu_\mathrm{B}\beta_\mathrm{e}^2\gamma_\mathrm{e}^2,
\label{eq:Esyn}
\end{equation}
and
\begin{equation}
\dot{E}_\mathrm{ic}=\frac{4}{3}\sigma_\mathrm{T}cu_\mathrm{rad}\beta_\mathrm{e}^2\gamma_\mathrm{e}^2,
\label{eq:Eic}
\end{equation}
\citep[e.g.][]{1979rpa..book.....R} where $\sigma_\mathrm{T}$ is the Thomson cross section. 
The energy loss rates are proportional to the energy densities, $u_\mathrm{B}$ and $u_\mathrm{rad}$, of the magnetic field and the seed photons, which are described later. 
The electron velocity $\beta_\mathrm{e}$ and the Lorentz factor $\gamma_\mathrm{e}$ are expressed in terms of the corresponding electron momentum $p_\mathrm{e}$ as follows,
\begin{equation}
\beta_\mathrm{e}=\frac{p_\mathrm{e}}{\sqrt{m_\mathrm{e}^2c^2+p_\mathrm{e}^2}},
\end{equation}
and
\begin{equation}
\gamma_\mathrm{e}=\sqrt{1+p_\mathrm{e}^2/(m_\mathrm{e}^2c^2)}.
\end{equation}
As we will see below, the photospheric emission serves as a dominant seed photon source for inverse Compton cooling. 
Thus, the seed photon temperature is of the order of $1$ eV. 
On the other hand, the injection momentum of electrons is typically $\sim 30m_\mathrm{e}c$ (see Section \ref{sec:electron_spectrum}). 
Therefore, the energy of most seed photons in the rest frame of non-thermal electrons is much smaller than the electron rest energy $m_\mathrm{e}c^2$, allowing us to neglect several processes reducing the efficiency of inverse Compton cooling, such as the Klein-Nishina suppression and the electron recoil effect. 

The electrons in the shell can also cool according to the expansion of the shell. 
The adiabatic momentum loss rate is given by
\begin{equation}
\dot{p}_\mathrm{ad}=\frac{p_\mathrm{e}}{3}\frac{\dot{V}}{V},
\end{equation}
where $V$ and $\dot{V}$ are the volume of the shell and its expansion rate. 
The temporal evolutions of these quantities are also obtained from the semi-analytic model. 

The governing equation (\ref{eq:momentum_equation}) is numerically solved by a simple upwind scheme with first-order implicit time integration. 
In the following calculations, the distributions of non-thermal electrons accelerated at the forward and reverse shock fronts are separately treated.

\subsection{Synchrotron spectrum}
The shock dissipation generates random magnetic field via some magnetohydrodynamics and/or plasma collective effects. 
In a similar way to the energy density of electrons, we use a parameter $\epsilon_\mathrm{B}$ describing the fraction of the magnetic field energy density to the dissipated shock energy. 
Thus, using the downstream internal energy densities $u_\mathrm{fs}$ and $u_\mathrm{rs}$ for the forward and reverse shocks, the corresponding magnetic energy densities are $u_\mathrm{B,fs}=\epsilon_\mathrm{B}u_\mathrm{fs}$ and $u_\mathrm{B,rs}=\epsilon_\mathrm{B}u_\mathrm{rs}$. 
These magnetic energy densities are used to evaluate the synchrotron energy loss rate, Equation (\ref{eq:Esyn}). 
The magnetic field strengths are given by 
\begin{equation}
B_\mathrm{fs}=(8\pi u_\mathrm{B,fs})^{1/2}=(8\pi \epsilon_\mathrm{B}u_\mathrm{fs})^{1/2},
\end{equation}
and
\begin{equation}
B_\mathrm{rs}=(8\pi u_\mathrm{B,rs})^{1/2}=(8\pi \epsilon_\mathrm{B}u_\mathrm{rs})^{1/2}.
\end{equation} 

For a given electron momentum distribution and a magnetic field strength, the synchrotron emissivity per unit frequency is calculated by the following formula, 
\begin{equation}
j_{\nu,\mathrm{syn}}=\frac{1}{4\pi V}\int P_{\nu,\mathrm{syn}}(\gamma_\mathrm{e})\frac{dN}{dp_\mathrm{e}}\mathrm{d}p_\mathrm{e}. 
\end{equation}
The synchrotron power per unit frequency $P_{\nu,\mathrm{syn}}(\gamma_\mathrm{e})$ as a function of electron Lorentz factor $\gamma_\mathrm{e}$ and frequency $\nu$ is described in Appendix \ref{sec:synchrotron}. 
At low frequencies, synchrotron emission suffers from absorption by its inverse process. 
The synchrotron self-absorption coefficient is given by  
\begin{equation}
\alpha_{\nu,\mathrm{syn}}=\frac{c^2}{8\pi V\nu^2}
\int\frac{\partial}{\partial p_\mathrm{e}}
\left[
p_\mathrm{e}\gamma_\mathrm{e}P_{\nu,\mathrm{syn}}(\gamma_\mathrm{e})
\right]
\frac{1}{p_\mathrm{e}^2}\frac{dN_\mathrm{e}}{dp_\mathrm{e}}dp_\mathrm{e}.
\end{equation}
Using these quantities and assuming that the emitting region is geometrically thin, the synchrotron intensity is obtained as follows,
\begin{equation}
I_\mathrm{syn}(\nu)=\frac{j_{\nu,\mathrm{syn}}}{\alpha_{\nu,\mathrm{syn}}}(1-e^{-\tau_{\nu,\mathrm{syn}}}),
\end{equation}
\citep[e.g.][]{1979rpa..book.....R}, where $\tau_{\nu,\mathrm{syn}}$ is the corresponding optical depth. 

\subsection{Inverse Compton spectrum}
We consider the photospheric emission from the ejecta as the dominant source of seed photons for inverse Compton emission. 
The radiation energy density corresponding to the photospheric luminosity $L_\mathrm{ph}$ is 
\begin{equation}
u_\mathrm{rad}=\frac{L_\mathrm{ph}}{4\pi cR_\mathrm{sh}^2},
\end{equation}
at the shell $r=R_\mathrm{sh}$, which is used to evaluate the inverse Compton energy loss rate, Equation (\ref{eq:Eic}). 
We assume that the photospheric emission is well represented by a blackbody spectrum with the colour temperature identical with $T_\mathrm{eff}$. 
Therefore, the photon spectrum is given by
\begin{equation}
\begin{aligned}
I_\mathrm{sn}(\nu)=&\frac{2u_\mathrm{rad}}{c^2a_\mathrm{r}T_\mathrm{eff}^4}
\frac{h\nu^3}{e^{h\nu/k_\mathrm{B}T_\mathrm{eff}}-1}
\\=&
\frac{L_\mathrm{sn}}{2\pi c^3a_\mathrm{r}T_\mathrm{eff}^4R_\mathrm{sh}^2}
\frac{h\nu^3}{e^{h\nu/k_\mathrm{B}T_\mathrm{eff}}-1},
\end{aligned}
\end{equation}
where $a_\mathrm{r}$ and $k_\mathrm{B}$ are the radiation constant and the Boltzmann constant. 

We also consider synchrotron emission as the other source of seed photons. 
We obtain the total intensity of seed photons by adding those of the photospheric emission and the synchrotron emission, $I_\mathrm{seed}(\nu)=I_\mathrm{sn}(\nu)+I_\mathrm{syn}(\nu)$. 
The spectrum $I_\mathrm{ic}(\nu)$ of the inverse Compton emission is calculated by convolving the seed photon spectrum, the electron distribution, and the redistribution function of Compton scattering $\Delta G$ as follows,
\begin{equation}
I_\mathrm{ic}(\nu)=\int _{p_\mathrm{min}}^{p_\mathrm{max}}
\Delta G
I_\mathrm{seed}(\nu')\frac{dN}{dp_\mathrm{e}}d\nu' dp_\mathrm{e},
\end{equation}
where the function $\Delta G$ is described in Section \ref{sec:compton}. 
\section{Results}\label{sec:results}
In this section, we show light curves and spectra calculated by the method described above. 
In all the calculations below, the microphysics parameters are assumed to be $p=3$, $\epsilon_\mathrm{e}=0.1$, and $\epsilon_\mathrm{B}=0.02$. 
We note that the electron spectral index $p=3$ is widely used for stripped-envelope CCSNe \citep[e.g.,][]{2006ApJ...651..381C}. 
The parameter $\epsilon_\mathrm{e}$ of the order of $0.1$ is also used in radio light curve modellings of highly energetic SNe including relativistic SNe \citep[e.g.,][]{2010Natur.463..513S,2015MNRAS.448..417B,2015ApJ...805..164N}, while the values of $\epsilon_\mathrm{B}$ show a variety depending on the radio brightness \citep[e.g.,][for GRB afterglows]{2014ApJ...785...29S}. 
We have determined the value of $\epsilon_\mathrm{B}$ so that the radio light curves of highly energetic SNe are reproduced by our fiducial model (see below). 

The ejecta mass and energy are also fixed to be $M_\mathrm{ej}=10M_\odot$ and $E_\mathrm{ej}=10^{52}$ erg, while we examine how light curves at different frequencies depend on the other parameters, $t_\mathrm{sd}$, $n$, and $A_\star$. 
Hereafter, the model with $t_\mathrm{sd}=10^{6}$ s, $n=6$, and $A_\star=1.0$ is called the fiducial model. 

\subsection{Photospheric emission}\label{sec:photospheric_emission}
First, we show theoretical light curves for photospheric emission from the ejecta powered by the central energy injection. 
In particular, we focus on the dependence of the light curves on the assumed spin down time. 

\begin{figure}
\begin{center}
\includegraphics[scale=0.55,bb=0 0 432 576]{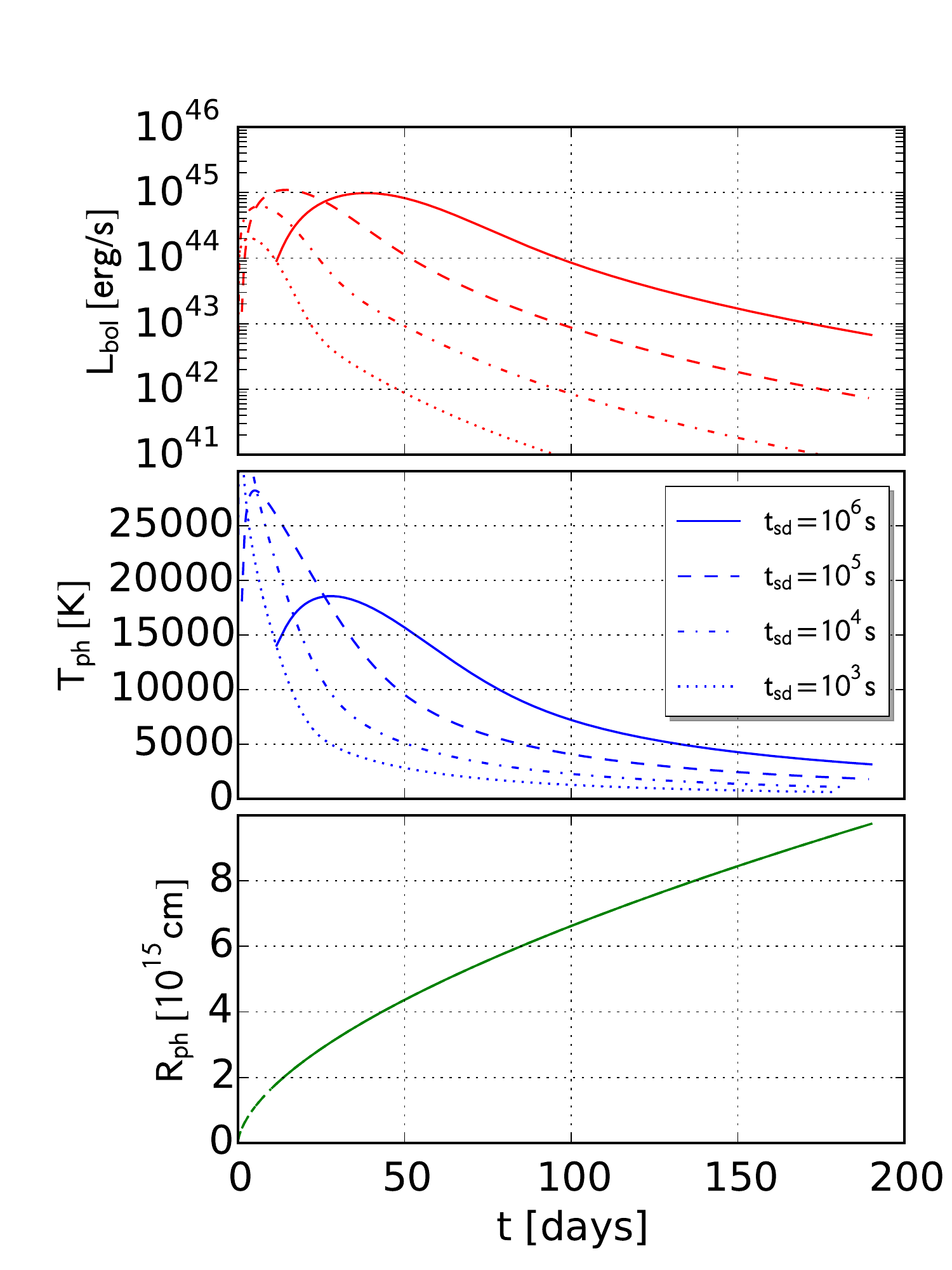}
\caption{Temporal evolutions of the photospheric luminosity (upper panel), the effective temperature (middle panel), and the photospheric radius (lower panel). 
Models with different spin down times are plotted in each panel. 
The solid, dashed, dash-dotted and dotted lines represent models with $t_\mathrm{sd}=10^{6}$, $10^{5}$, $10^{4}$, and $10^{3}$ s. }
\label{fig:Lsn}
\end{center}
\end{figure}

\begin{figure*}
\begin{center}
\includegraphics[scale=0.55,bb=0 0 850 566]{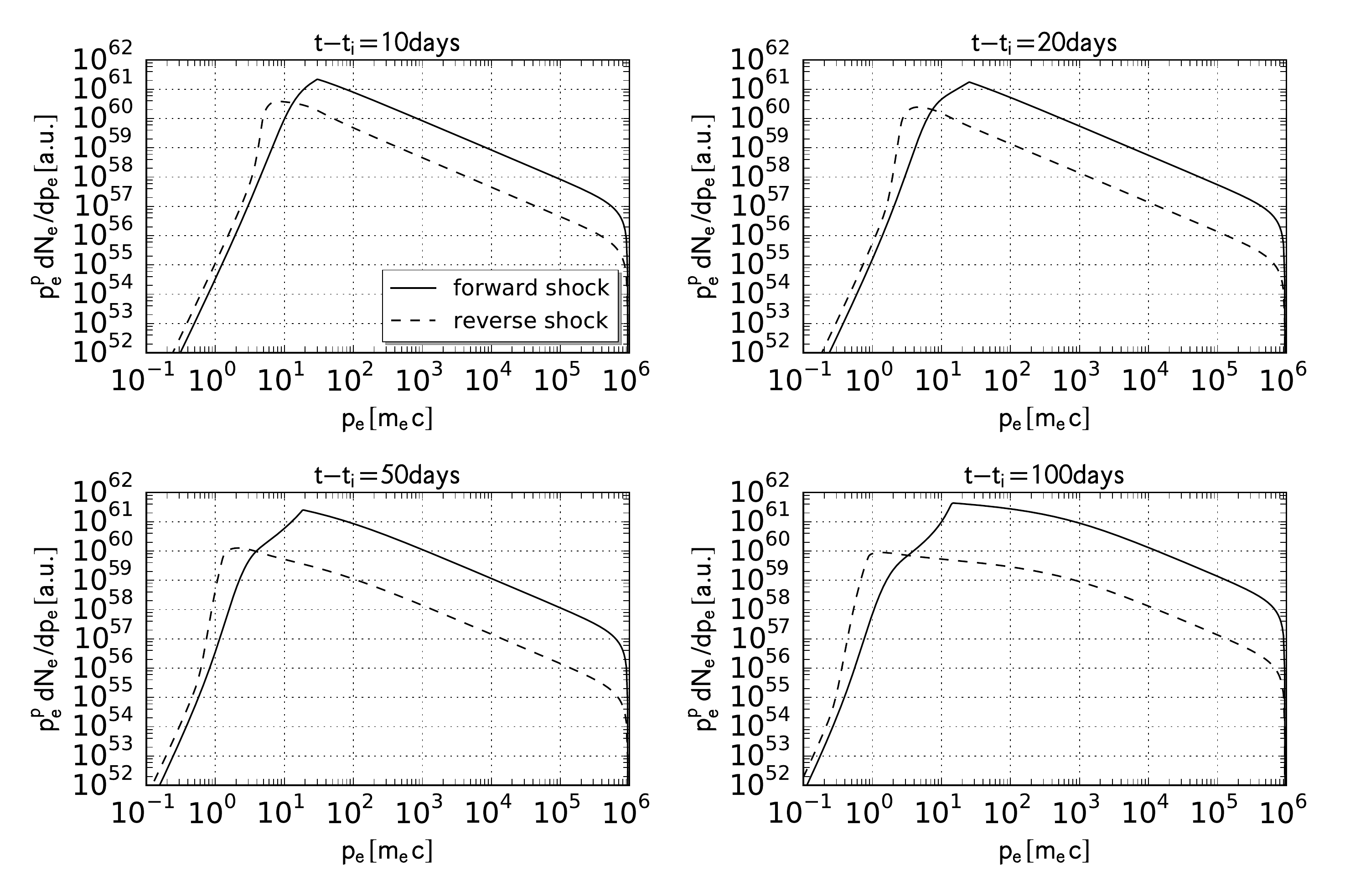}
\caption{Electron momentum distributions multiplied by $p_\mathrm{e}^p$ ($p_\mathrm{e}^pdN/dp_\mathrm{e}$) at $t-t_\mathrm{i}=10$ (top left), $20$ (top right), $50$ (bottom left), and $100$ (bottom right) days. 
In each panel, the solid and dashed curves correspond to the contributions from the forward and reverse shocks. 
}
\label{fig:espec}
\end{center}
\end{figure*}

Figure \ref{fig:Lsn} shows the temporal evolutions of the photospheric luminosity, the effective temperature at the photosphere, and the photospheric radius. 
The duration and the peak luminosity of the bolometric light curve become shorter and less luminous for shorter spin down times. 
Therefore models with shorter $t_\mathrm{sd}$ result in small radiated energies. 
In these models, the energy injection is terminated at early stages of the evolution of the ejecta. 
The timescale of the energy injection is only a small fraction of the timescale at which the ejecta becomes transparent to thermal photons. 
Therefore, the injected energy suffers from significant adiabatic loss before escaping into the interstellar space as radiation, leading to low radiative efficiencies. 
In other words, the ratio of the spin down time $t_\mathrm{sd}$ to the diffusion time $t_\mathrm{d}$ plays a critical role in determining bolometric light curves of central engine-powered SNe \citep{2010ApJ...717..245K,2015MNRAS.454.3311M,2015MNRAS.452.3869N,2017ApJ...850...55N}. 
The model with $t_\mathrm{sd}=10^{6}$ s can reproduce the timescale and the peak bolometric luminosity of SLSNe-I. 
On the other hand, the model with $t_\mathrm{sd}=10^{3}$ s exhibit a shorter timescale and a lower peak luminosity than SLSNe-I. 
Such models may be relevant to SNe associated with GRBs. 
The over-luminous SN 2011kl associated with the ultra-long GRB 111209A exhibited a fast evolving light curve with a timescale of $10$--$20$ days and a peak luminosity of $\sim 3\times 10^{43}$ erg s$^{-1}$. 
The timescale is similar to that of the model with $t_\mathrm{sd}=10^{3}$ s, while the peak luminosity is smaller by a factor of $\sim 6$. 
The discrepancy could be resolved by adjusting the free parameters or considering jet-like energy injection rather than a quasi-spherical wind. 

In our model, the photospheric radius at time $t$ is solely determined by the density structure of the ejecta. 
Thus, the temporal evolutions of $R_\mathrm{ph}$ for different models are exactly same as shown in the bottom panel of Figure \ref{fig:Lsn}. 
The entire ejecta become transparent at $t\simeq 190$ days, after which both the photospheric radius and the effective temperature cannot be well defined and thus we stop calculations. 

The temporal evolution of the photospheric radius and the spectral evolution have been obtained for the nearby slowly evolving SLSN-I 2015bn \citep{2016ApJ...826...39N,2016ApJ...828L..18N,2017ApJ...835...13J}. 
The measured photospheric radius of SLSNe-I 2015bn rose until $\sim 50$ days after the optical maximum and then started declining after reaching the maximum value of $R_\mathrm{ph}\simeq 10^{16}$ cm. 
The decline of the bolometric light curve becomes steeper around $200$--$300$ days after the optical maximum, which is accompanied by a gradual transition from a continuum-dominated spectrum to a nebular one. 
This epoch of the transition is roughly consistent with the time at which the ejecta used in our model become transparent at $t\simeq 190 $ days. 
However, the temporal behavior of the photospheric radius should be calculated by a more sophisticated treatment of radiative transfer and the internal structure of the ejecta. 
These values would be sensitive to the opacity for thermal photons and possibly to three-dimensional ejecta structure. 
In reality, the ejecta structure would be patchy due to the ``shredding'' by gas flows at high Lorentz factors penetrating the entire ejecta \citep{2003ApJ...589..871A,2011MNRAS.411.2054L,2017MNRAS.466.2633S}. 
Therefore, when taking the multi-dimensional effect into account, optical depths corresponding to radial rays with different directions could differ from each other. 
This indicates that more thorough investigations including effects of three-dimensional ejecta structure and sophisticated radiative transfer in the ejecta are required.

\subsection{Electron momentum distribution}\label{sec:electron_spectrum}

Figure \ref{fig:espec} shows the electron momentum distributions at $t-t_\mathrm{i}=10$, $20$, $50$, and $100$ days. 
The plotted distributions are multiplied by $p_\mathrm{e}^p$ ($p_\mathrm{e}^pdN/dp_\mathrm{e}$) so that the spectrum of the injected electrons appears to be flat. 
In this model, electrons accelerated at the forward shock front are more abundant and have higher average energy than those at the reverse shock, which reflects the large energy dissipation rate at the forward shock (see Figure \ref{fig:dynamics}). 
These electrons behind the forward shock thus predominantly contribute to the non-thermal emission. 
The distributions are divided into two segments separated by a peak. 
The peak momentum corresponds to the minimum injected momentum. 
As Figure \ref{fig:dynamics} shows, the shock velocity is $v_\mathrm{sh}\simeq 0.3$--$0.4$c at several 10 days. 
The kinetic energy density of the flow is roughly given by $\rho v_\mathrm{sh}^2$, and a considerable fraction of this energy is supposed to dissipate at the shock front, $u_\mathrm{int}\sim \rho v_\mathrm{sh}^2$. 
Therefore, the average Lorentz factor of electrons is roughly estimated from Equation (\ref{eq:electron_gamma}),
\begin{equation}
\hat{\gamma}=\epsilon_\mathrm{e}\frac{Am_\mathrm{u}v_\mathrm{sh}^2}{Zm_\mathrm{e}c^2}\simeq 33
\left(\frac{\epsilon_\mathrm{e}}{0.1}\right)
\left(\frac{v_\mathrm{sh}/c}{0.3}\right)^2
\left(\frac{Z/A}{0.5}\right)^{-1},
\end{equation}
for the forward shock, which agrees with the peak momenta shown in Figure \ref{fig:espec}. 
The lower peak momenta for the reverse shock are due to small shock velocities relative to the unshocked ejecta velocities. 

At higher energies, the distributions are well represented by a power-law function with an index $-(p+1)$, $dN/dp_\mathrm{p}\propto p_\mathrm{e}^{-(p+1)}$. 
This indicates that the fast cooling regime \citep{1998ApJ...497L..17S,2001ApJ...548..787S} is realized at earlier epochs. 
This is because thermal photons abundantly produced by the photospheric emission can efficiently cool non-thermal electrons via inverse Compton scattering. 
At later epochs, e.g., the distribution at $t-t_\mathrm{i}=100$ days, a relatively flat distribution at $p_\mathrm{e}/(m_\mathrm{e}c)=20$--$100$ indicates that injected electrons with lower energies remain uncooled because of the declining photospheric luminosity. 
At lower energies than the peak, on the other hand, the electron momentum distribution shows a hard spectrum, which is composed of electrons having lost most of their energies.

\subsection{Radio light curve}\label{sec:radio_lc}
\begin{figure}
\begin{center}
\includegraphics[scale=0.50,bb=0 0 453 680]{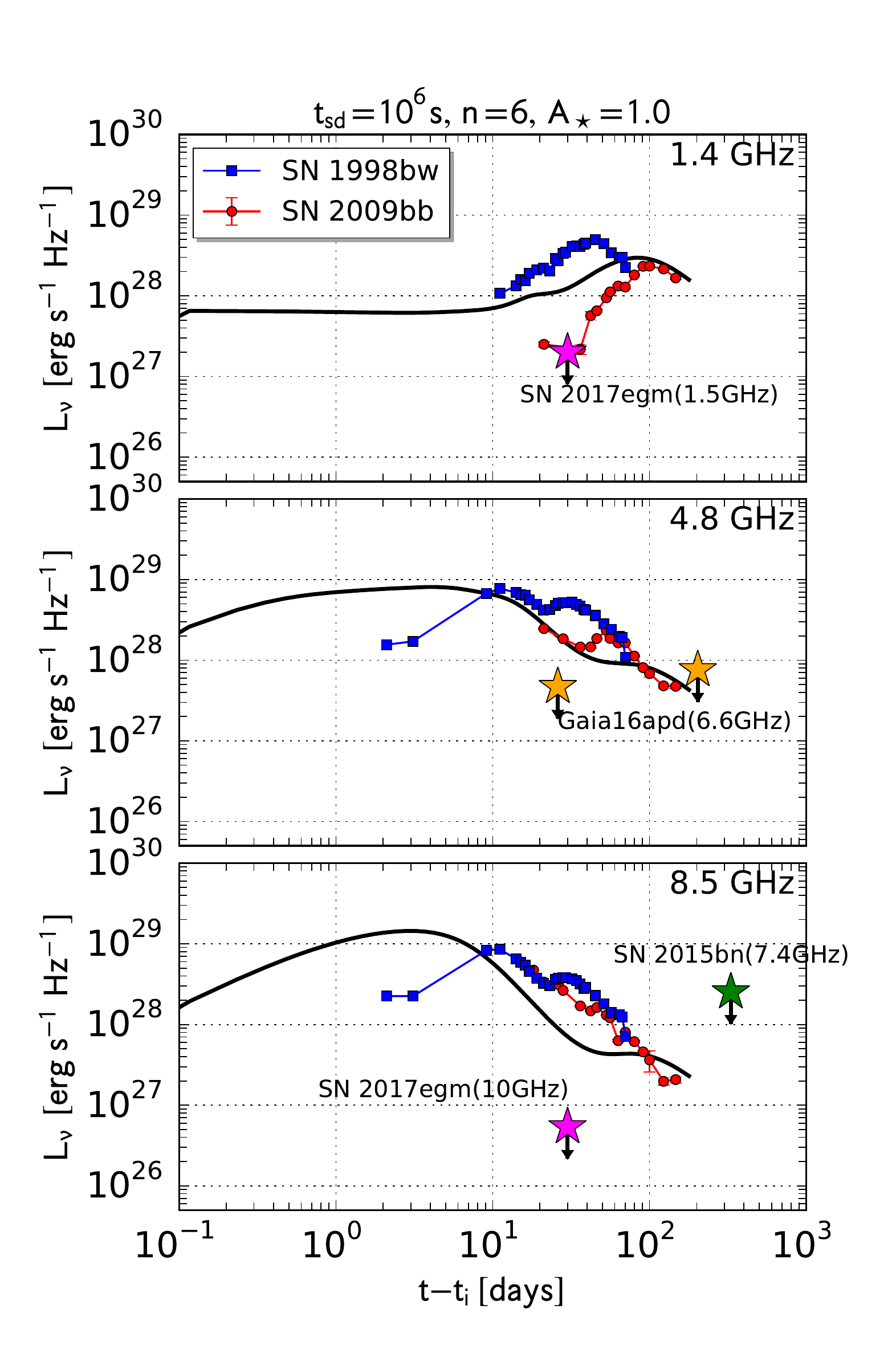}
\caption{Radio light curves of the non-thermal emission model for $1.4$ (top), $4.8$ (middle), and $8.5$ (bottom) Hz. 
The solid lines show the radio light curve calculated by the fiducial model with $t_\mathrm{sd}=10^{6}$ s, $n=6$ and $A_\star=1$. 
The microscopic free parameters are set to be $p=3.0$, $\epsilon_\mathrm{e}=0.1$, and $\epsilon_\mathrm{B}=0.02$. 
For comparison, light curves of radio-loud SNe Ic-BL, 1998bw (blue square) and 2009bb (red circle) at the corresponding frequency, are also plotted. 
The star marks with arrows represent upper limits obtained by radio observations for two SLSNe-I. 
The green star in the bottom panel represents the upper limit for SLSN-I 2015bn at $7.4$ GHz, the orange stars in the middle panel represent the upper limits for Gaia16apd at $6.6$ GHz,  and the magenta stars in the top and bottom panels are those for SLSN-I 2017egm at $1.5$ and $10$ GHz. }
\label{fig:radio}
\end{center}
\end{figure}

Figure \ref{fig:radio} shows the radio light curves at different frequencies, $1.4$, $4.8$, and $8.5$ GHz, calculated by our fiducial model with $t_\mathrm{sd}=10^{6}$ s, $n=6$, and $A_\star=1$. 
Because of the hot bubble breakout and the subsequent acceleration of the outermost layers of the ejecta, central engine powered SNe give rise to bright radio emission especially at early epochs. 
The radio light curves of our fiducial model suggest that the radio luminosity exhibits a peak around $\sim 5$--$10$ days for $\nu\simeq 5$--$10$ GHz, while the peak at $\nu\simeq 1$ GHz appears around $50$--$100$ days. 

These features are worth comparing with radio-loud SNe. 
In Figure \ref{fig:radio}, we plot the radio light curves of SNe 1998bw and 2009bb for comparison. 
SN 1998bw was a widely known SN Ic-BL associated with GRB 980425 \citep{1998Natur.395..663K,1998Natur.395..670G}. 
SN 2009bb was the SN Ic-BL whose properties were remarkably similar to GRB-associated SNe, but lacking any signature of gamma-ray emission \citep{2010Natur.463..513S}. 
As shown in Figure \ref{fig:radio}, their radio luminosities were similar to each other. 
For SN 1998bw, the peak of the light curve was successfully observed at $1.4$, $4.8$ and $8.5$ GHz thanks to early observations triggered by the gamma-ray detection. 
The peak was earlier at higher frequencies as is the case for radio emission from normal CCSNe interacting with their CSM \citep[e.g.][]{2016arXiv161207459C}. 
For SN 2009bb, the peaks at higher frequencies, $\nu=4.8$ and $8.5$ Hz, were probably missed, while the peak at $1.4$ GHz was successfully observed. 
The decline rates of the luminosities per unit frequency after the peak are similar for both events. 
For SN 2009bb, the presence of an ultra-relativistic jet is unlikely because of the absence of emission indicating off-axis jet. 
The radio emission is explained by trans-relativistic supernova ejecta (\citealt{2010Natur.463..513S}, see also \citealt{2015ApJ...805..164N}).

We also plot the upper limits obtained by radio observations of SN 2015bn \citep{2016ApJ...826...39N}, Gaia16apd \citep{2018ApJ...856...56C}, and 2017egm \citep{2018ApJ...853...57B} in Figure \ref{fig:radio}. 
We should note that the frequency bands for SN 2015bn ($7.4$ GHz), Gaia16apd ($6.6$ GHz), and 2017egm ($1.5$ and $10$ GHz) are slightly different from the theoretical light curve and SNe 1998bw and 2009bb ($1.5$ and $8.5$ GHz). 
However, the spectral energy distributions of the synchrotron emission (see Figure \ref{fig:sed}) suggest that the radio luminosities at the corresponding frequency bands are similar to the theoretical light curve shown in Figure \ref{fig:radio} within a factor of a few.

\begin{figure*}
\begin{center}
\includegraphics[scale=0.38,bb=0 0 1359 680]{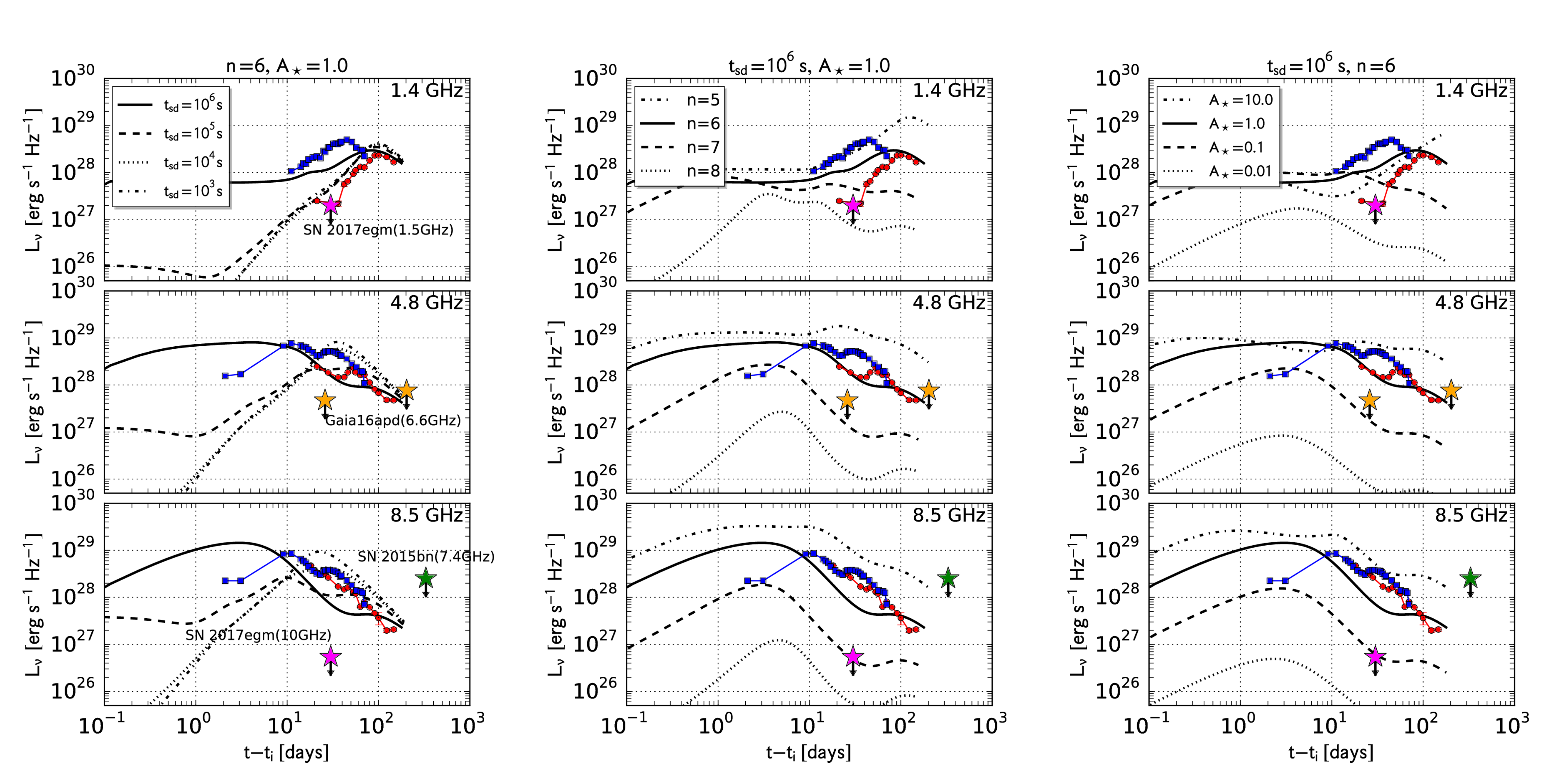}
\caption{Dependence of the radio light curves on the spin down time $t_\mathrm{sd}$ (left column), the power-law exponent $n$ of the density profile $n$ (middle column), and the CSM density $A_\star$ (right column). 
In the left panels, we compare the models with $t_\mathrm{sd}=10^{6}$(solid), $10^{5}$(dashed), $10^{4}$(dotted), and $10^{3}$(dash-dotted) s. 
The models with $n=5$ (dash-dotted) , $6$ (solid), $7$ (dashed), and $8$ (dotted) are shown in the middle panels. 
The right panels represent the models with $A_\star=10.0$ (dash-dotted), $1.0$ (solid), $0.1$ (dashed), and $0.01$ (dotted). 
The other free parameters are set to the same one as the fiducial model in Figure \ref{fig:radio}. }
\label{fig:radio2}
\end{center}
\end{figure*}

In Figure \ref{fig:radio2}, we show how the radio light curves depend on the free parameters, the spin down time $t_\mathrm{sd}$, the power-law exponent $n$ of the density profile, and the CSM density $A_\star$. 
We first focus on the effect of the spin down time. 
As is seen in the left column of Figure \ref{fig:radio2}, the models with longer $t_\mathrm{sd}$ exhibit bright radio emission in early epochs but are less luminous at later epochs than those with shorter $t_\mathrm{sd}$. 
The power-law exponent more significantly affects the radio light curve than the spin down time, since it determines how much fraction of the kinetic energy is distributed in the outermost layers interacting with the CSM. 
For shallower density slopes (smaller $n$), more energy is available in the outermost layer to produce non-thermal electrons, giving rise to brighter synchrotron emission. 
This trend of brighter radio luminosities for shallower density slopes is seen in the middle column of Figure \ref{fig:radio2}. 
The increase in the CSM density makes the emission brighter because a dense CSM can efficiently dissipate the kinetic energy of the ejecta. 

The radio light curves of our fiducial model in Figure \ref{fig:radio} show good agreement with SNe 1998bw and 2009bb. 
The radio luminosities at $8.5$, $4.8$, $1.4$ GHz show their peaks around $t\simeq 5$, $7$, and $50$ days. 
After the peak, the model luminosity steadily declines at a rate similar to those observed for SNe 1998bw and 2009bb. 
The radio non-detection of SN 2015bn is consistent with most of the models. 
Since the upper limit of the radio luminosity is smaller than the peak luminosity of the fiducial model, earlier radio observations might have detected radio emission from the SLSN. 
On the other hand, the fiducial model disagrees with the radio observations of SN 2017egm at $\sim30$ days after the detection. 
Thanks to the proximity of the SLSN, the tightest constraint for radio emission of SLSNe-I is obtained. 
Most of theoretical radio light curves exceed the upper limit for a range of the parameters. 
In order to explain the radio emission from SN 2017egm, either a steep density gradient $(n\la 8)$ or a small CSM density $A_\star\la 0.01$ is required. 
The upper limit for Gaia 16apd at 26 days after the detection also places a meaningful constraint on the radio luminosity. 
In similar ways to SN 2017egm, steep density gradients and/or small CSM densities are required to reconcile the disagreement. 
We will further discuss theoretical interpretations of the current radio constraints in Section \ref{sec:conclusion}. 

\subsection{X-ray light curve}
\begin{figure*}
\begin{center}
\includegraphics[scale=0.50,bb=0 0 850 850]{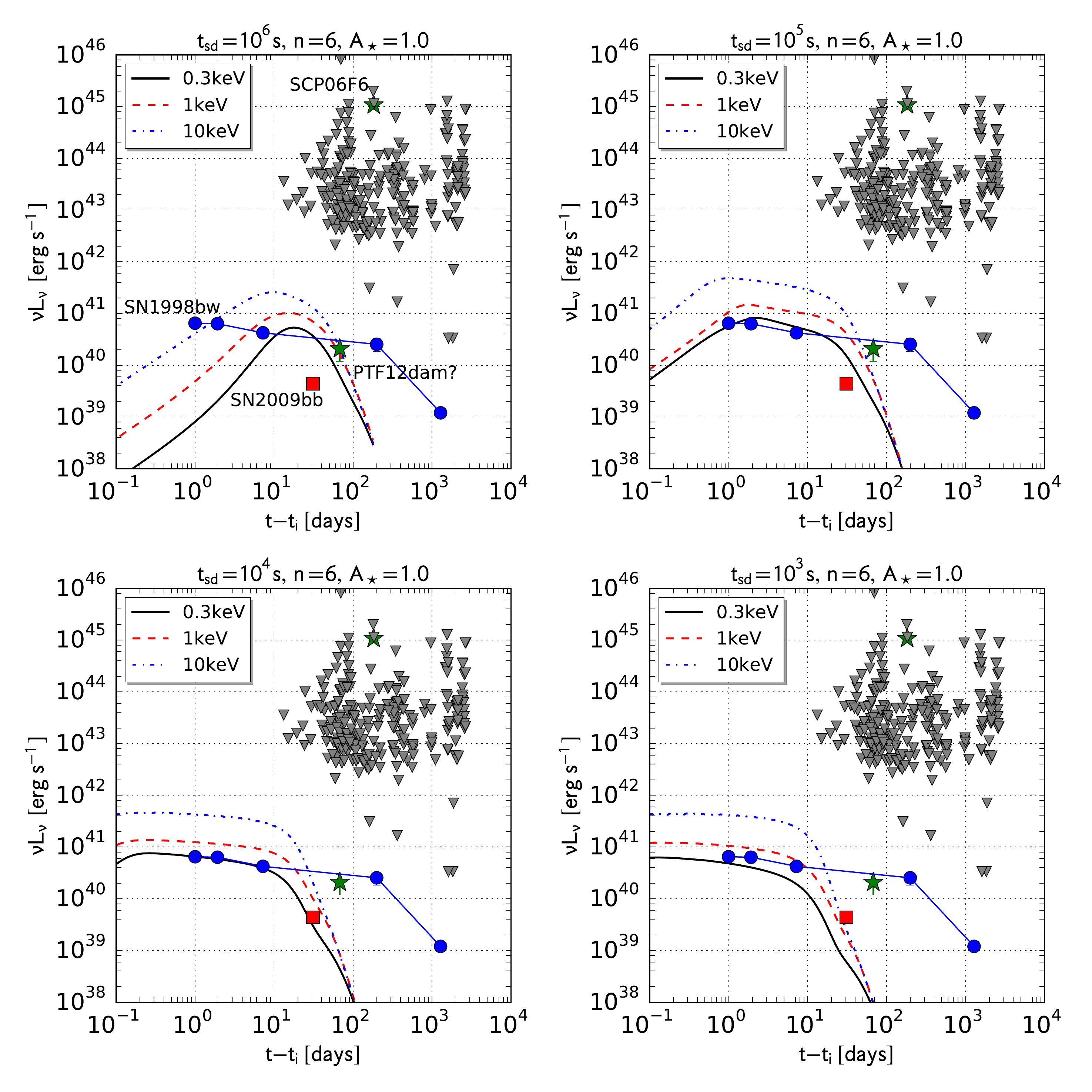}
\caption{X-ray light curves of models with $t_\mathrm{sd}=10^{6}$ (top left), $10^{5}$ (top right), $10^{4}$ (bottom left), and $10^{3}$ (bottom right) s. 
Theoretical $\nu L_\nu$ light curves are plotted for $0.3$, $1.0$, and $10$ keV and $0.3$--$10$ X-ray light curves of some SNe are compared. 
Green stars represent the (possible) X-ray detection of SLSNe SCP 06F6 and PTF12dam. 
We also plot SNe Ic-BL 1998bw (blue circles) and  2009bb (red square) for comparison. 
SLSNe upper limits are plotted as gray triangles. 
Plotted data adopted from \protect\cite{2017arXiv170405865M} for SLSNe, \protect\cite{2000ApJ...536..778P} and \protect\cite{2004ApJ...608..872K} for SN 1998bw, and \protect\cite{2010Natur.463..513S} for SN 2009bb.
}
\label{fig:xray}
\end{center}
\end{figure*}

Figure \ref{fig:xray} shows the X-ray light curves of the inverse Compton emission. 
The curves in the panels of Figure \ref{fig:xray} present the temporal evolution of $\nu L_\nu$ at $h\nu=0.3$, $1.0$, and $10$ keV for models with $t_\mathrm{sd}=10^{6}$, $10^{5}$, $10^{4}$, and $10^{3}$ s. 
The other parameters are fixed as $n=6$ and $A_\star=1.0$. 
The luminosity of the inverse Compton emission from the ejecta can reach $\nu L_\nu\simeq 10^{41-42}$ erg s$^{-1}$, which is brighter than most of normal CCSNe. 
The luminous X-ray emission is owing to the presence of the mildly relativistic ejecta and the optical photons abundantly provided by the photospheric emission. 
We plot the currently available upper limits for SLSNe-I (including SN 2015bn) and $0.3$--$10$keV X-ray luminosities of SCP 06F6  and PTF12dam (we used the data compiled by \cite{2017arXiv170405865M}). 
The luminous X-ray emission associated with SCP 06F6 have generated a lot of discussion on its origin \citep{2009ApJ...697L.129G,2013ApJ...771..136L,2014MNRAS.437..703M}. 
However, the X-ray luminosity is well above most of the upper limits obtained for SLSNe-I so far (Figure \ref{fig:xray}), leading to the consensus that such luminous X-ray emission is not common among SLSNe-I. 
\cite{2017arXiv170405865M} found an X-ray source at the location of PTF12dam. 
However, as they mention in their paper, the X-ray source can also be explained by X-ray emission associated with the star-forming activity in the host galaxy with a relatively high star formation rate $\sim 5M_\odot$ yr$^{-1}$. 
Therefore, further observations are required to see whether the X-ray emission is certainly associated with PTF12dam or not. 
In Figure \ref{fig:xray}, we also plot the X-ray light curves of the SNe Ic-BL 1998bw \citep{2000ApJ...536..778P,2004ApJ...608..872K} and 2009bb \citep{2010Natur.463..513S} for comparison. 

We first focus on our fiducial model in the top left panel of Figure \ref{fig:xray}. 
The theoretical light curve is below most of the upper limits placed for the other SLSNe, suggesting that deeper observations are needed to further constrain the central engine scenario for SLSNe-I. 
The X-ray luminosity of PTF12dam agrees with the theoretical value, but it may have to be treated as an upper limit because of the reason described above. 
The theoretical light curves exhibit their peaks around 10-30 days. 
Since the luminosity of inverse Compton emission is proportional to the product of the seed photon energy density and the energy of non-thermal electrons, the light curve is determined by the convolution of the bolometric light curve of the photospheric emission and the steadily declining energy dissipation rate at the shock front (see Figure \ref{fig:dynamics}). 
Thus, the peak in the X-ray light curve slightly precedes the optical maximum. 

The theoretical light curves of the other models exhibit similar X-ray luminosities but earlier peaks for shorter spin down times. 
This is because the optical maximum shifts earlier for shorter $t_\mathrm{sd}$ as we have described in Section \ref{sec:photospheric_emission}. 
For models with shorter $t_\mathrm{sd}$, the light curve exhibits a plateau rather than a peak and then the luminosity declines. 
This feature is similar to the X-ray light curve of SN 1998bw, although the observed light curve shows a longer flat part. 
Although only a single data point is available, SN 2009bb also show similar X-ray luminosity, which agrees with the declining theoretical light curve at $\sim$30 days. 
One caveat on this comparison is that the corresponding theoretical bolometric luminosities of the photospheric emission (Figure \ref{fig:Lsn}) are brighter than those of SN 1998bw and SN 2009bb. 
This discrepancy indicates that we should explore appropriate parameters satisfying both optical and X-ray observational constraints and/or an improved treatment of the photospheric emission with multi-colour radiation transfer and other sources of seed photons would be required. 
We leave such improvements to future work. 

As in the case of radio light curves, increasing the CSM density $A_\star$ makes the X-ray emission more luminous. 
Since the theoretical X-ray light curve of the fiducial model with $A_\star=1.0$  in the upper left panel of Figure \ref{fig:xray} roughly matches the X-ray flux of PTF12dam, the CSM density much larger than this value would predict too bright X-ray emission. 
This can place an upper limit on the CSM density by treating the X-ray flux as an upper limit. 
The adopted value $A_\star=1.0$ corresponds to a steady wind at a mass-loss rate of $\dot{M}=10^{-5}\ M_\odot$ yr$^{-1}$ for a wind velocity $10^3$ km s$^{-1}$. 
Therefore, mass-loss rates much larger than this value is unlikely. 
\cite{2017arXiv170405865M} have already constrained the CSM density by using the X-ray upper limit and reached a similar conclusion, $\dot{M}<2\times 10^{-5}\ M_\odot$ yr$^{-1}$. 
\subsection{Broad-band spectral energy distribution}
\begin{figure*}
\begin{center}
\includegraphics[scale=0.55,bb=0 0 850 566]{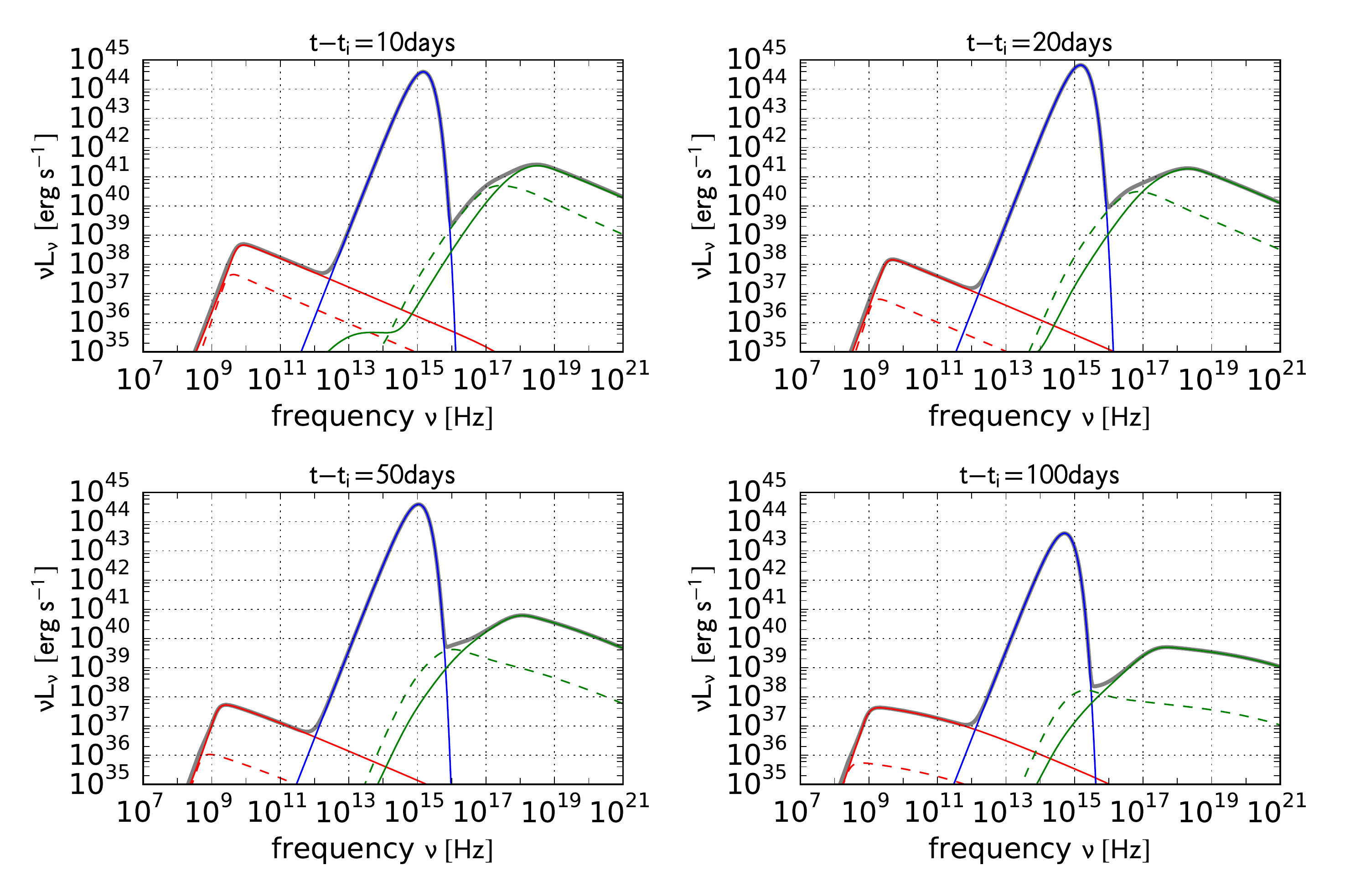}
\caption{Spectral energy distribution at $t-t_\mathrm{i}=10$ (top left), $20$ (top right), $50$ (bottom left), and $100$ (bottom right) days. 
In each panel, the red, blue, and green curves, whose peaks are at around $10^9$--$10^{10}$, $10^{15}$, and $10^{18}$ Hz, represent the synchrotron, photospheric, and inverse Compton components. 
The solid and dashed curves correspond to the contributions from the reverse and forward shocks, while the gray thick curve is the sum of all the contributions.  
}
\label{fig:sed}
\end{center}
\end{figure*}

Finally, we present spectral energy distributions at several epochs in Figure \ref{fig:sed}. 
The spectral energy distributions at different epochs look similar to each other, and qualitatively similar to those of normal CCSNe interacting with their CSM. 
Each distribution is composed of the synchrotron, photospheric, and inverse Compton components, whose peaks are located around $10^9$--$10^{10}$, $10^{15}$, and $10^{18}$ Hz. 
The peak in the radio energy range divides the synchrotron component into optically thick (lower frequencies) and thin (higher frequencies) regimes. 
The peak shifts toward lower frequencies with time because the non-thermal electrons in the shocked region gradually become transparent to radio waves with lower frequencies. 
This temporal shift of the radio peak frequency creates the peak in the radio light curve described in Section \ref{sec:radio_lc}
The flux of the optically thick synchrotron emission follows a power-law function of $\nu$, $L_\nu\propto \nu^{5/2}$ \citep[e.g.][]{1979rpa..book.....R}.
On the other hand, the spectral slope of the optically thin synchrotron emission depends on the power-law exponent $p$ of the electron energy spectrum. 
As we have described in Section \ref{sec:electron_spectrum}, the fast cooling regime can apply, $L_\nu\propto \nu^{-p/2}=\nu^{-1.5}$. 

Since the inverse Compton spectrum is the convolution of the energy spectra of non-thermal electrons and seed photons, the peak frequency of the inverse Compton component is determined by the peak in the electron momentum distribution and the effective temperature of the photospheric emission. 
As we have described in Section \ref{sec:electron_spectrum}, the injection energy of non-thermal electrons is $\hat{\gamma}\simeq 30$--$40$. 
The inverse Compton scattering by electrons with this Lorentz factor increases the photon energy by a factor $\hat{\gamma}^2$, 
\begin{equation}
\nu_\mathrm{ic}\simeq \frac{3\hat{\gamma}^2k_\mathrm{B}T_\mathrm{eff}}{h}=10^{18}\mathrm{Hz}
\left(\frac{\hat{\gamma}}{40}\right)^2
\left(\frac{T_\mathrm{eff}}{10^4\mathrm{K}}\right),
\end{equation}
which explains the X-ray peak in the spectral energy distribution. 
The spectrum at frequencies higher than the X-ray peak frequency is well represented by a power-law distribution, whose exponent depends on the electron energy spectrum. 
The spectral index is same as the optically thin synchrotron emission, $L_\nu\propto \nu^{-p/2}=\nu^{-1.5}$. 

\section{Discussion and conclusions}\label{sec:conclusion}
In this study, we have calculated broad-band emission from supernova ejecta powered by a central engine based on the picture revealed by the recent two-dimensional special relativistic hydrodynamic simulation \citep{2017MNRAS.466.2633S}. 
In the hydrodynamic simulation, the outermost layers of the ejecta are efficiently accelerated owing to a hot gas emerging from the central region of the ejecta and thus the maximum velocity of the ejecta can be mildly relativistic. 
While the ejecta emit thermal photons by radiative diffusion, the outermost layers colliding with a CSM create the forward and reverse shocks propagating in the CSM and ejecta, respectively. 
We model the photospheric emission from the ejecta by using the Arnett-type one-zone model for photon diffusion throughout the ejecta and at the same time we determine the photospheric radius and the effective temperature from the ejecta model. 
Furthermore, we calculated non-thermal emission from the shocked gas by using the semi-analytic model for the propagation of the forward and reverse shocks \citep{2017ApJ...834...32S}. 

We found that non-thermal electrons produced in the shocked region can give rise to bright radio and X-ray emission via synchrotron and inverse Compton processes. 
When we adopt commonly assumed values for microphysics parameters, $\epsilon_\mathrm{e}$, $\epsilon_\mathrm{B}$, and $p$, and the CSM density corresponding to a steady mass-loss rate of $\dot{M}=10^{-5}\ M_\odot$ yr$^{-1}$ and a constant wind velocity of $10^3$ km s$^{-1}$, the theoretical radio light curves of central engine powered SNe well agree with those of radio-loud SNe Ic-BL, such as SNe 1998bw and 2009bb, consistent with the idea that they also harbour a central engine. 

\subsection{Radio and X-ray emission from SLSNe-I as a probe of a central engine}
Our results suggest that SLSNe-I can also give rise to radio synchrotron emission with similar fluxes to radio-loud SNe Ic-BL. 
This could also be explained naturally if the SLSNe-I and SNe Ic-BL would be certainly linked to each other as suggested by similarities in spectra of SNe Ic-BL and SLSNe-I \citep{2010ApJ...724L..16P,2017ApJ...835...13J,2016ApJ...828L..18N,2017ApJ...845...85L}. 
As shown in Figure \ref{fig:radio2}, the radio brightness of the central-engine powered supernova ejecta highly depends on the density profile of the ejecta and the CSM density. 
Our previous hydrodynamics simulation \citep{2017MNRAS.466.2633S} suggests that if an SN harbours a sufficiently energetic central engine to produce the hot bubble breakout, the impact of the blowout would create an ejecta component travelling at relativistic speeds. 
This situation corresponds to models with shallow density gradient ($n=5$ and $6$), which produce bright radio emission. 
Therefore, radio observations of SLSNe-I may probe whether the SLSN-I have experienced the hot bubble breakout.  

The radio upper limit for SN 2015bn \citep{2016ApJ...826...39N} is well above the theoretical light curves of our fiducial model (Figure \ref{fig:radio}). 
If radio emission of SN 2015bn was as bright as SN 1998bw and 2009bb, it could have been detected by observations conducted earlier (at several 10 days). 
On the other hand, for the recently discovered SLSN-I 2017egm and Gaia16apd, tighter upper limits are available. 
These upper limits rule out most of the models with shallow density gradients and dense CSM densities. 

X-ray emission can also assess the presence of relativistic ejecta. 
The predicted X-ray fluxes are much lower than most of the currently available upper limits for SLSNe-I. 
The theoretical models cannot explain the X-ray luminosity of SCP 06F6, indicating a different origin for the unusually bright X-ray emission. 
Our fiducial model agrees with the X-ray luminosity of PTF12dam, although it requires further observations to confirm the association of the X-ray source with PTF12dam. 
Furthermore, models with short spin down times well explain the X-ray emission from the SNe Ic-BL, 1998bw and 2009bb. 
Although X-ray observations of SLSNe-I are currently not so constraining as radio observations, future X-ray observations can also be used as a powerful tool for investigating the outermost ejecta structure of SLSNe-I and other extraordinary SNe. 

We consider two possibilities to interpret the radio non-detections. 
First, some SLSNe may not experience the hot bubble breakout because only a small amount of additional energy is injected from the central engine. 
Although the kinetic energy of the supernova explosion preceding the central energy injection is not known, if we assume a typical kinetic energy of $10^{51}$ erg, the additional energy required to produce the hot bubble breakout ranges from a few $10^{51}$ to $10^{52}$ erg, depending on the density structure of the supernova ejecta \citep{2017MNRAS.466.2633S,2017ApJ...845..139B}. 
The light curve fitting of multi-colour optical data of SN 2017egm \citep{2017ApJ...845L...8N} infers a relatively small kinetic energy, $1$--$2\times 10^{51}$ erg, for this nearby event. 
The kinetic energy of Gaia16apd is estimated to be $3.69^{+1.38}_{-0.59}\times 10^{51}$ erg \citep{2017ApJ...850...55N}. 
The estimated kinetic energy of the ejecta also depend on the density structure of the freely expanding ejecta assumed in the light curve fitting model. 
Although there are uncertainties in the critical energy for the hot bubble breakout and the kinetic energy of the supernova ejecta, the injected energy may be smaller than the critical energy and thus relativistic ejecta may not be produced. 
\cite{2017ApJ...850...55N} also performed light curve fitting to other SNSNe-I. 
According to their results, the ejecta mass and the kinetic energy of SLSNe-I are distributed in the range of $2.2$--$12.9\ M_\odot$ and $(1.9$-$9.8)\times10^{51}$ erg, respectively. 
Thus, the ejecta mass and energy assumed in our model are close to the upper end of the distributions. 
On the other hand, the spin down time $t_\mathrm{sd}=10^6$ s employed in our fiducial model is typical among the SLSNe. 
This may suggest the possibility that not all SLSNe-I experience the hot bubble breakout with bright radio emission.

The other possibility is that the density slope of the outermost layer of SLSNe ejecta is not so shallow as predicted by the hydrodynamic simulation by \cite{2017MNRAS.466.2633S}. 
Although the hydrodynamic simulation suggests the presence of the mildly relativistic ejecta, it is still unclear whether the relativistic component is realized when correctly taking into account coupling between gas and radiation. 
In the hydrodynamics simulation without radiative transfer, gas and radiation are assumed to be strongly coupled, which enables the efficient acceleration of the outermost layers by radiation pressure. 
However, in reality, gas and radiation may be coupled only weakly in the outermost layers. 
In other words, radiation in the outermost layer may simply escape into the surrounding space rather than accelerating gas in the layer, leading to smaller kinetic energy and maximum velocity of the ejecta in the homologous expansion stage. 
This may be especially true for SLSNe-I, because they require spin down timescales comparable to the diffusion time of thermal photons in the ejecta.

\subsection{SN explosions with a central engine}
From the results of our broad-band light curve modelling combined with the dynamical evolution of SNe with central energy sources revealed by \cite{2017MNRAS.466.2633S}, we can speculate the following scenario for SNe with central energy sources. 

The most important factor is the total amount of the injected energy. 
For an injected energy exceeding a critical value depending on the original ejecta structure, the ejecta are significantly affected by the energy injection. 
Even when the additional energy is deposited as thermal energy, a quasi-spherical relativistic wind would soon be created around the centre and start pushing the ejecta. 
The hot bubble breakout and the associated energy redistribution throughout the ejecta potentially produce mildly relativistic ejecta with a shallow density gradient. 

The presence of relativistic ejecta depends on the timescale of the energy injection compared with the diffusion timescale of the ejecta. 
For central energy injection with much shorter duration than the diffusion timescale, the injected energy would predominantly be converted to the kinetic energy of the ejecta via adiabatic expansion rather than escaping as thermal photons. 
This case likely produces supernova ejecta with a relatively large kinetic energy, which may be observed as SNe Ic-BL. 
On the other hand, for energy injection timescales comparable to or longer than the diffusion timescale, the injected energy can easily escape into interstellar space as thermal photons, giving rise to bright thermal emission. 
This may correspond to SLSNe-I. 
In terms of their radio and X-ray properties, SLSNe-I produced in such a way can be divided into two classes. 
One is the population harbouring sufficiently energetic central engine to produce the hot bubble breakout and thus they are radio-loud. 
The other is the population whose central engine can give rise to bright optical emission but is not accompanied by relativistic ejecta. 

\subsection{Other remarks}
Finally, we mention the following two remarks on the broad-band light curve modelling. 
We should note that the expected radio and X-ray luminosities highly depend on the CSM density and the density profile of the supernova ejecta. 
The circumstellar environments of SLSNe-I are poorly known. 
They may explode in relatively clean environments, making non-thermal emission weak. 
The radio and X-ray light curve modelling of SLSNe-I significantly suffer from these uncertainties. 

Another potential caveat is that the non-thermal emission could also arise from the wind nebula of the nascent neutron star. 
Recent theoretical modellings of non-thermal emission from the wind nebula embedded in spherical supernova ejecta suggest that the non-thermal emission start leaking the dense supernova ejecta after $\sim 100$ days \citep{2013MNRAS.432.3228K,2014MNRAS.437..703M,2015ApJ...805...82M,2016ApJ...818...94K,2018MNRAS.474..573O}. 
Although how early X-ray and radio emission starts leaking depends on the multi-dimensional density structure of the supernova ejecta, non-thermal emission from the wind nebula would basically be preceded by that from the shock interaction at the ejecta-CSM interface. 
Therefore, radio and X-ray detections at early epochs likely indicate non-thermal emission from the blast wave driven by the supernova ejecta. 

\appendix
\section{Radiative processes}\label{sec:radiative_processes}
In this section, we summarize several formulae for radiative processes used in our non-thermal emission model. 
\subsection{Synchrotron emission}\label{sec:synchrotron}
Synchrotron radiation power per unit frequency by a single electron with a Lorentz factor $\gamma_\mathrm{e}$ is given by
\begin{equation}
P_{\nu,\mathrm{syn}}(\gamma_\mathrm{e})=
\frac{\sqrt{3}e^3B\sin \theta_\mathrm{p}}{3m_\mathrm{e}c^2}F(\nu/\nu_\mathrm{syn}),
\end{equation}
with the synchrotron frequency $\nu_\mathrm{syn}$ being
\begin{equation}
\nu_\mathrm{syn}=\frac{3\gamma_\mathrm{e}^2eB}{2m_\mathrm{e}c}\sin\theta_\mathrm{p},
\end{equation}
where $e$ is the elementary charge and the pitch angle $\theta_\mathrm{p}$ specifies the angle between the electron orbit and the magnetic field line. 
In this paper, we set the pitch angle to be
\begin{equation}
\sin\theta_\mathrm{p}=\sqrt{\frac{2}{3}},
\end{equation}
in order for the power integrated with respect to the solid angle matches the synchrotron energy loss rate, Equation (\ref{eq:Esyn}). 
The frequency dependence of the synchrotron power per unit frequency is determined by the function $F(x)$, which is given by the following form,
\begin{equation}
F(x)=\int^\infty_xK_{5/3}(y)dy,
\end{equation}
where $K_{5/3}(y)$ is the modified Bessel function of the second kind with an order $5/3$
This function can be numerically evaluated in a straight forward way. 

\subsection{Inverse Compton emission}\label{sec:compton}
The interaction between an electron and a photon has long been considered. 
The redistribution function of Compton scattering gives the distribution of an outgoing photon in the energy space for an incoming photon and is calculated by carefully integrating the differential cross section of the process \citep[e.g.][]{1968PhRv..167.1159J,1973erh..book.....P,1986JQSRT..36..273K,1990MNRAS.245..453C}. 

In the same way as previous work \citep[e.g.,][]{2009ApJ...698..293V}, we obtain the redistribution function fully taking account relativistic effects by integrating the covariant form of the different cross section (or the invariant scattering amplitude). 
The result is described as follows,
\begin{equation}
\begin{aligned}
\Delta G(E_\mathrm{e},E_{\gamma}',E_\gamma)=&
\sum_{i=0}^4\left[G_i(w_+,E_\mathrm{e},E_{\gamma}',E_\gamma)-G_i(w_-,E_\mathrm{e},E_{\gamma}',E_\gamma)\right]
\\&+G_5(w_+,E_\mathrm{e},E_{\gamma}',E_\gamma)+G_5(w_-,E_\mathrm{e},E_{\gamma}',E_\gamma),
\end{aligned}
\end{equation}
where $E_\mathrm{e}$ is the energy of the incoming electron and $E_\gamma'$ and $E_\gamma$ are the energies of incoming and scattered photons. 
The functions $G_i\ (i=0\mathrm{-}5)$ are given by
\begin{equation}
G_0(w,E_\mathrm{e},E_{\gamma}',E_\gamma)=
-\frac{2}{E_\gamma E_\gamma'}\left[(E_\gamma'-E_\gamma)^2+\frac{4E_\gamma E_\gamma'}{1+w^2}\right]^{1/2}
,
\end{equation}
\begin{equation}
G_1(w,E_\mathrm{e},E_{\gamma}',E_\gamma)=
-\frac{2m^2_\mathrm{e}c^4}{E_\gamma^2E_\gamma'^2}\sqrt{E^2_\mathrm{e}+m^2_\mathrm{e}c^4w^2}
,
\end{equation}
\begin{equation}
G_2(w,E_\mathrm{e},E_{\gamma}',E_\gamma)=
-\frac{2\sqrt{E^2_\mathrm{e}+m^2_\mathrm{e}c^4w^2}}{(E^2_\mathrm{e}-m^2_\mathrm{e}c^4)(1+w^2)}
,
\end{equation}
\begin{equation}
G_3(w,E_\mathrm{e},E_{\gamma}',E_\gamma)=
-\frac{2m^2_\mathrm{e}c^4}{E_\gamma E_\gamma'}\frac{E^2_\mathrm{e}}{(E^2_\mathrm{e}-m^2_\mathrm{e}c^4)\sqrt{E^2_\mathrm{e}+m^2_\mathrm{e}c^4w^2}}
,
\end{equation}
\begin{equation}
\begin{aligned}
G_4(w,E_\mathrm{e},E_{\gamma}',E_\gamma)=&
\frac{2m^2_\mathrm{e}c^4}{(E_\mathrm{e}^2-m_\mathrm{e}^2c^4)^{3/2}}
\left(1+\frac{2E^2_\mathrm{e}-m^2_\mathrm{e}c^4}{E_\gamma E_\gamma'}\right)
\\&\times\mathrm{Arctanh}
\left(\sqrt{\frac{E_\mathrm{e}^2+m_\mathrm{e}^2c^4w^2}{E_\mathrm{e}^2-m_\mathrm{e}^2c^4}}\right)
,
\end{aligned}
\end{equation}
and
\begin{equation}
G_5(w,E_\mathrm{e},E_{\gamma}',E_\gamma)=
-\frac{m^2_\mathrm{e}c^4(E_\gamma+E_\gamma')}{E_\gamma^2 E_\gamma'^2}\frac{E_\mathrm{e}}{\sqrt{E^2_\mathrm{e}+m^2_
\mathrm{e}c^4w^2}}
.
\end{equation}
The quantity $w$ in in these functions is related to the cosine of the scattering angle $\mu$,
\begin{equation}
w^2=\frac{1+\mu}{1-\mu},
\end{equation}
and the values of the scattering angle $\mu$ corresponding to $w_\pm$ are obtained by solving the following quartic equation,
\begin{equation}
\begin{aligned}
E_\gamma^2 E_\gamma'^2(1-\mu)^2&
+2E_\gamma E_\gamma'
\left[E_\mathrm{e}(E_\gamma'+E_\mathrm{e}-E_\gamma)-m_\mathrm{e}^2c^4\right](1-\mu)\\
&+m_\mathrm{e}^2c^4(E_\gamma-E_\gamma')^2=0.
\end{aligned}
\end{equation}
The larger and smaller values of $w$ corresponding to the solutions are denoted by $w_+$ and $w_-$, respectively.

\section*{Acknowledgements}
We thank the anonymous referee for his or her constructive comments. 
This research was supported by MEXT as a Priority Issue on the post-K computer (Elucidation of the Fundamental Laws and Evolution of the Universe) and the Joint Institute for Computational Fundamental Science (JICFuS). 
K.M. acknowledges support by the Japan Society for the Promotion of Science (JSPS) KAKENHI grant 17H02864.








\bsp	
\label{lastpage}
\end{document}